\documentclass[11pt, a4paper, twoside]{article}

\usepackage[a4paper, top=3cm, bottom=3cm, left=3cm, right=3cm]{geometry}
\providecommand{\keywords}[1]
{
  \small	
  \textbf{\textit{Keywords---}} #1
}
\usepackage[round]{natbib}
\usepackage{hyperref}

\usepackage{graphicx, xcolor, ae, fancyvrb}
\definecolor{Red}{rgb}{0.5,0,0}
\definecolor{Blue}{rgb}{0,0,0.5}
\hypersetup{%
    hyperindex = {true},
    colorlinks = {true},
    linktocpage = {true},
    plainpages = {false},
    linkcolor = {Red},
    citecolor = {Blue},
    urlcolor = {Red},
    pdfstartview = {Fit},
    pdfpagemode = {UseOutlines},
    pdfview = {XYZ null null null}
  }

\usepackage{caption}
\usepackage{amsmath, amssymb, bbm, caption, subcaption, multicol}

\usepackage{tikz}
\usetikzlibrary{arrows, matrix}

\usepackage{algorithm}
\usepackage{algpseudocode}
\newcommand{\reals}{\mathbb{R}}
\newcommand{\proc}{\texttt{Proc}}
\newcommand{\one}{\mathbbm{1}}
\newcommand{\B}{\mathcal{B}}
\newcommand{\prob}{\mathbb{P}}
\newcommand{\N}{\mathcal{N}}
\newcommand{\Dir}{\mathrm{Dir}}
\newcommand{\per}{\mathrm{per}}

\newcommand{\Z}{\mathbb{Z}}

\newcommand{\Logistic}{\mathrm{Logistic}}

\DeclareMathOperator*{\argmax}{arg\,max}

\author{Dennis Christensen\\University of Oslo\\Norwegian Defence Research Establishment\\ \texttt{dennis.christensen@ffi.no} \and Per August Jarval Moen\\University of Oslo \\ \texttt{pamoen@math.uio.no}}
\DefineVerbatimEnvironment{Code}{Verbatim}{}
\DefineVerbatimEnvironment{CodeInput}{Verbatim}{fontshape=sl}
\DefineVerbatimEnvironment{CodeOutput}{Verbatim}{}
\newenvironment{CodeChunk}{}{}
\setkeys{Gin}{width=0.8\textwidth}

\makeatletter
\newcommand\code{\bgroup
  \catcode`\_ 12 % Make underscore a printable character
  \catcode`\$ 12 % Make dollar sign a printable character
  \ttfamily % Switch to typewriter (monospaced) font
  \@codex
}
\def\@codex#1{{\normalfont\ttfamily\hyphenchar\font=`\-#1}\egroup}
\makeatother

\title{\texttt{perms}: Likelihood-free estimation of marginal likelihoods for binary response data in Python and R}

\begin{document}
\maketitle

%% - \Abstract{} almost as usual
\begin{abstract}
In Bayesian statistics, the marginal likelihood (ML) is the key ingredient needed for model comparison and model averaging. Unfortunately, estimating MLs accurately is notoriously difficult, especially for models where posterior simulation is not possible. Recently, \citet{christensen2023inference} introduced the concept of permutation counting, which can accurately estimate MLs of models for exchangeable binary responses. Such data arise in a multitude of statistical problems, including binary classification, bioassay and sensitivity testing. Permutation counting is entirely likelihood-free and works for any model from which a random sample can be generated, including nonparametric models. Here we present \texttt{perms}, a package implementing permutation counting. As a result of extensive optimisation efforts, \texttt{perms} is computationally efficient and able to handle large data problems. It is available as both an R package and a Python library. A broad gallery of examples illustrating its usage is provided, which includes both standard parametric binary classification and novel applications of nonparametric models, such as changepoint analysis. We also cover the details of the implementation of perms and illustrate its computational speed via a simple simulation study.
\end{abstract}

%% - \Keywords{} with LaTeX markup, at least one required
%% - \Plainkeywords{} without LaTeX markup (if necessary)
%% - Should be comma-separated and in sentence case.

%% - \Address{} of at least one author
%% - May contain multiple affiliations for each author
%%   (in extra lines, separated by \emph{and}\\).
%% - May contain multiple authors for the same affiliation
%%   (in the same first line, separated by comma).

%% -- Introduction -------------------------------------------------------------

%% - In principle "as usual".
%% - But should typically have some discussion of both _software_ and _methods_.
%% - Use \proglang{}, \pkg{}, and \code{} markup throughout the manuscript.
%% - If such markup is in (sub)section titles, a plain text version has to be
%%   added as well.
%% - All software mentioned should be properly \cite-d.
%% - All abbreviations should be introduced.
%% - Unless the expansions of abbreviations are proper names (like "Journal
%%   of Statistical Software" above) they should be in sentence case (like
%%   "generalized linear models" below).

\section{Introduction}
\subsection{Bayesian binary response models}\label{sec:intro}
Binary responses arise naturally in a wide variety of statistical problems. One key example is that of binary classification, in which we estimate the probability that a given input should be classified with the label $y=1$ rather than $y=0$. Another example is that of bioassay problems, in which a series of experiments are performed at different dosage levels, and the researcher observes whether each experiment leads to a success ($y=1$) or failure ($y=0$). In industry, experiments of this kind are commonly referred to as sensitivity testing. For example, in the energetic materials industry, they are used to estimate the susceptibility of an explosive to ignite under various levels of physical impact \citep{christensen2023improved}.

The analysis of binary responses can be unified in a single framework. Let $\theta$ denote the parameters of the model. In the Bayesian setting, $\theta$ is treated as random, with a prior distribution $\theta\sim\pi(\theta)$. Depending on the model choice, $\theta$ could be a finite-dimensional vector (the parametric case) or an infinite-dimensional object like a probability distribution or a function (the nonparametric case). Given $\theta$, a vector $T = T(\theta) = (T_1, \dots, T_n)$ of thresholds is generated from a deterministic function of $\theta$, which possibly also depends on covariates. Further, a vector of latent variables $X = (X_1, \dots, X_n)$ is generated from some distribution $\pi(x\mid\theta)$, where we assume that the $X_i$ are conditionally i.i.d. given $\theta$. Finally, the binary responses $Y_i$ are modelled as indicator variables,
\begin{equation}\label{eq:indicatorY}
    Y_i = \one\{X_i \leq T_i\}.
\end{equation}
The above setup is highly general and covers most binary response data models, as the following list of examples indicates. In what follows, let $z_i \in \mathbb{R}^p$ be a fixed vector of covariates for $i=1, \ldots, n$.

\begin{itemize}
    \item \textbf{Bayesian logistic regression}. Let $T_i = \theta^\top z_i$, where the coefficient vector $\theta$ takes values in $\mathbb{R}^p$, and let $X_i\sim\Logistic(1)$, so that the cdf of $X_i$ is $\sigma(s) = 1/(1 + \exp(-s))$. Then
    $$\prob(Y_i = 1\mid \theta) = \prob(X_i \leq T_i\mid \theta) = \sigma(\theta^\top z_i),$$
    so we recover standard Bayesian logistic regression. Other binary generalised linear models like the probit and complementary log-log regression models can be recovered in the same fashion by adjusting the cdf of the $X_i$.
    
    \item \textbf{Gaussian process classification}. Let $T_i = \theta(z_i)$, where the random function $\theta:\reals^p \rightarrow \reals$ is distributed according to a Gaussian process, and let $X_i \sim \Logistic(1)$ like before. Then
    $$\prob(Y_i = 1\mid \theta) = \sigma(\theta(z_i)),$$
    so we recover binary classification with Gaussian processes. 

    \item \textbf{Parametric bioassay}. Suppose that the $T_i = t_i$ are fixed and known, and let $\pi_X(\cdot\mid\theta)$ be a probability distribution with cdf $F_\theta$. Then
    $$\prob(Y_i = 1\mid\theta) = F_\theta(t_i),$$
    a standard parametric bioassay setup.

    \item \textbf{Nonparametric bioassay}. Like before, suppose the $T_i = t_i$ are fixed and known, but now let $X_i\mid\theta\sim \theta$, where $\theta$ is a random probability distribution on the reals, generated by some process, $\theta\sim\proc$. For example, $\proc$ could be the Dirichlet process \citep{ferguson1973bayesian}. Then $\theta$ emits a random cdf $F$, and
    $$\prob(Y_i = 1\mid \theta) = F(t_i).$$
    Such models were first studied by \citet{antoniak1974mixtures}.
\end{itemize}

In Bayesian statistics, model selection and model averaging all depend on computing the marginal likelihood (ML),
$$\pi(y) = \prob(Y = y),$$
which is generally intractable. For binary responses, the ML is therefore needed for tasks like feature selection (a form of model selection), or ensembling over different models to increase predictive accuracy (model averaging). If the likelihood function is computable, then posterior inference is possible via Markov chain Monte Carlo (MCMC) methods, and MLs can be estimated from a posterior sample using e.g. bridge esimators \citep{meng1996simulating}. However, for nonparametric models, or any model in which the likelihood function is intractable, there is no standard recipe for estimating MLs.

Recently, \citet{christensen2023inference} proposed a new simulation technique based on permutation counting, which provides an unbiased and consistent estimator of the ML requiring only exchangeable binary response data. Unlike approaches based on MCMC, permutation counting is entirely likelihood-free and works for any model from which samples from the forward model can be generated. In particular, it works for virtually all binary Bayesian nonparametric models. The main idea behind permutation counting is to exploit the symmetries introduced by exchangeability, and then to correct for this exploitation by calculating appropriate weights, called permutation numbers. It turns out that these can be represented as permanents of a certain class of binary matrices, called \emph{block rectangular}. \citet{christensen2023inference} proved that these permanents can be calculated in polynomial time, and provided a Python implementation of this algorithm, which is able to handle small data problems ($n < 250$).

In this paper we introduce \texttt{perms}, which is both a  Python library and an R package for estimating marginal likelihoods via permutation counting. Several optimising steps have been taken to maximise the efficiency of the package. Among these are memory recycling and exploitation of the sparsity of a key data structure to minimise memory usage and overhead. Additionally, the core part of \texttt{perms} is written in the C programming language. Along with a substantial improvement in speed, our implementation is able tackle problems with data of sample size up to several thousands on a desktop computer.

The remainder of this paper is structured as follows. Section~\ref{sec:permutationcounting} establishes basic notation and introduces the idea of permutation counting. Section~\ref{sec:basicusage} covers basic usage of \texttt{perms}, both in Python and R, by considering a tractable toy problem. In Section~\ref{sec:marginal_likelihoods_perms} we show how \texttt{perms} can be used to estimate MLs in real problems by considering a tractable binary classification problem and a nonparametric bioassay problem. This section also explains how \texttt{perms} can handle bioassay type data (with repeated trials), saving the user from reformatting the data. %Section~\ref{sec:marginal_likelihoods_perms} shows how to estimate the ML in a standard binary classification setup, and compares the results with benchmark approaches. We also discuss how to h. 
Section~\ref{sec:computation} details how \texttt{perms} is implemented, and Section~\ref{sec:implementation_details} discusses the various steps taken to minimise computational cost. Section~\ref{sec:results} illustrates the speed gain of \texttt{perms} via simulations, comparing it to \citeauthor{christensen2023inference}'s original implementation. Finally, in Section~\ref{sec:examplecp}, we use \texttt{perms} to solve a nonparametric changepoint analysis problem, illustrating how \texttt{perms} can be used to solve more complicated tasks. 

For reproducibilty, replication code is attached as supplementary material, which includes all code displayed in the paper, as well as code used for the simulation studies. The reported execution times of all code in this paper stem from running the code on a MacBook Pro (2021 model) running MacOS 12.3 with an Apple M1 CPU and 16GB DDR4 RAM.

\subsection{Permutation counting}\label{sec:permutationcounting}
We now set up the basic problem and present the notion of permutation counting.%, as introduced by \citet{christensen2023inference}
 From \eqref{eq:indicatorY}, we see that observing $Y_i = 1$ is equivalent to observing the event $\{X_i \leq T_i\}$. Thinking of our data in terms of sets, let us define the (random) set $\B_T = \B_{T,1}\times\cdots\times\B_{T,n}\subseteq\reals^n$ by
$$\B_{T,i} = \begin{cases} (-\infty, T_i] & \mathrm{ if}\,y_i=1, \\ (T_i, \infty) & \mathrm{ if}\,y_i=0, \end{cases}$$
for $i=1, \dots, n$. We see that observing $\{Y=y\}$ is equivalent to  observing $\{X\in\B_T\}$. In particular, the ML can be written as $\prob(Y = y) = \prob(X \in \B_T)$.

\begin{figure}
        \centering
        \begin{tikzpicture}[scale=2, xscale=1]

        % Frame
        \draw[thick, black] (0, 2.12) -- (0,0) -- (5, 0) -- (5, 2.12);

        % Left censor
        \draw[thick, black, -o] (0, 2*.5) node[left] {$\B_{T,1}$} -- (2, 2*.5);
        \draw[thick, black, -o] (0, 2*.75) node[left] {$\B_{T,2}, \B_{T,3}$} -- (3, 2*.75);
        \draw[thick, black, -o] (0, 2*.75 + .12) -- (3, 2*.75 + .12);
        \draw[thick, black, -o] (0, 2*1 + .12) node[left] {$\B_{T,4}$} -- (4, 2*1 + .12);
    
        % Right censor
        \draw[thick, black, -*] (5, 2*.25) node[right] {$\B_{T,5}$} -- (1, 2*.25);
        \draw[thick, black, -*] (5, 2*.5) node[right] {$\B_{T,6}$} -- (2, 2*.5);
        \draw[thick, black, -*] (5, 2*.75) node[right] {$\B_{T,7}$} -- (3, 2*.75);

        % x
        \draw[thick, black] (.5, -.1) node[below] {$X_5$} -- (.5, .1);
        \draw[thick, black] (1.5, -.1) node[below] {$X_2, X_3, X_7$} -- (1.5, .1);
        \draw[thick, black] (2.5, -.1) node[below] {$X_1$} -- (2.5, .1);
        \draw[thick, black] (3.5, -.1) node[below] {$X_6$} -- (3.5, .1);
        \draw[thick, black] (4.5, -.1) node[below] {$X_4$} -- (4.5, .1);
        
        \end{tikzpicture}
        \caption{\label{fig:favourite} Visual representation of the random sets $\B_{T,i}$, stacked vertically for clarity, and a random sample $X = (X_1, \dots, X_n)$, for $n=7$.}
\end{figure}
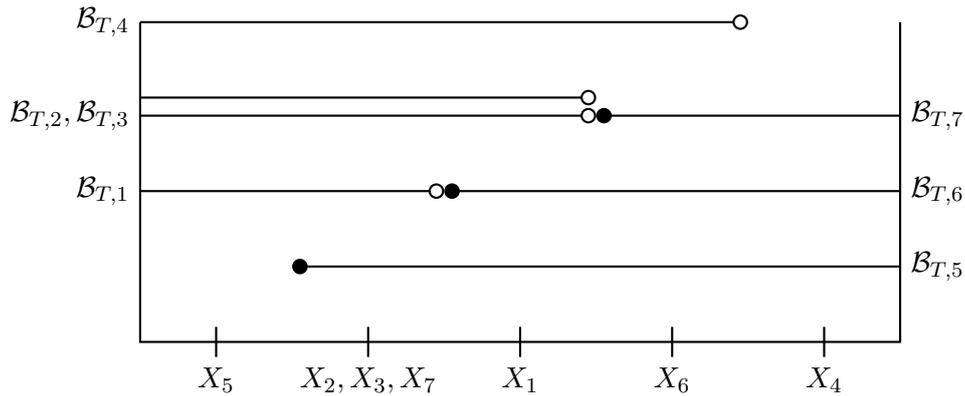

The key to permutation counting is to consider what happens when we permute the indices of $X$. Let $S_n$ denote the group of $n$-permutations. For any permutation $\sigma\in S_n$, we write 
$$\sigma(X) = (X_{\sigma(1)}, \dots, X_{\sigma(n)}).$$
Now, given $X\in\reals^n$, the \emph{permutation number} $w(X;\B_T)$ of $X$ (with respect to $\B_T$) is the number of permutations $\sigma$ such that $\sigma(X)\in\B_T$. That is,
\begin{equation}
    w(X; \B_T) = \#\{\sigma\in S_n : \sigma(X) \in \B_T\}.
\end{equation}
We note that $0 \leq w(X; \B_T) \leq n!$. Assuming for now that they can be computed efficiently, these permutation numbers naturally yield an unbiased and consistent estimator of the ML $\prob(X\in\B_T)$, namely
\begin{equation}
    \hat\pi_S = \frac1S\sum_{s=1}^S\frac1{n!}w(X^{(s)};\B_{T^{(s)}})\label{eq:logMLest},
\end{equation}
where $X^{(s)}\sim \pi(x\mid\theta^{(s)})$, $T^{(s)} = T(\theta^{(s)})$ and $\theta^{(s)}\sim\pi(\theta)$, independently, for $s=1, \dots, S$. Note that \eqref{eq:logMLest} does not require any evaluations of the likelihood function, and is entirely nonparametric. In Figure~\ref{fig:favourite}, we see a visual representation of the sets $\B_{T,i}$ for $n=7$, along with a single simulated sample $X$. This example is taken from \citet{christensen2023inference}. We see that $X\notin\B_T$, but for example $\sigma(X) = (X_5, X_2, X_3, X_7, X_1, X_6, X_4)\in\B$, so $w(X;\B_T)>0$ even though $X\notin\B_T$.

Permutation counting can also be used for posterior estimation. Indeed, let $h(\theta)$ be some quantity we wish to estimate. Then we have the following consistent (yet biased) estimator of the posterior mean $\mathbb{E}[h(\theta)\mid X\in\B_T]$,
\begin{equation}
    \hat h_S = \frac{\sum_{s=1}^Sh(\theta^{(s)}) w(X^{(s)};\B_{T^{(s)}})}{\sum_{s=1}^S w(X^{(s)}; \B_{T^{(s)}})}.
\end{equation}

Thus, the key problem at hand is to compute the permutation numbers $w(X;\B_T)$ efficiently. To do so, we need to introduce the notion of \emph{permanents}. Given an $n\times n$ matrix $A = (a_{ij})$, its permanent $\per\, A$ is defined as
    \begin{equation}\label{eq:defpermanent}
    \per\,A = \sum_{\sigma\in S_n}\prod_{i=1}^n a_{i, \sigma(i)}.
\end{equation}
It is not difficult to show that we can write $w(X;\B_T) = \per\,A$, where the $n\times n$ matrix $A=(a_{ij})$ is defined by
\begin{equation}\label{eq:matrixeq}
    a_{ij} = \begin{cases}1 & \mathrm{if}\,X_i\in\B_{T,j}, \\ 0 & \mathrm{otherwise}.\end{cases}
\end{equation}
After ordering the columns, the matrix $A$ corresponding to Figure~\ref{fig:favourite} is given by
\begin{equation}\label{eq:favourite}A_1 = \begin{tikzpicture}[baseline={(0,0)}]
            \matrix[matrix of math nodes, left delimiter=(, right delimiter=)]{ 
                1 & 1 & 1 & 1 & 0 & 0 & 0 \\
                1 & 1 & 1 & 1 & 1 & 0 & 0 \\
                1 & 1 & 1 & 1 & 1 & 0 & 0 \\
                1 & 1 & 1 & 1 & 1 & 1 & 0 \\
                0 & 1 & 1 & 1 & 1 & 1 & 1 \\
                0 & 0 & 0 & 0 & 1 & 1 & 1 \\
                0 & 0 & 0 & 0 & 0 & 1 & 1 \\
                };
            \draw (-3.5*.412, 3.5*.46) rectangle (-2.5*.412,-.5*.46);
            \draw (-2.5*.412, 3.5*.46) rectangle (.5*.412, -1.5*.46);
            \draw (.5*.412, 2.5*.46) rectangle (1.5*.412, -2.5*.46);
            \draw (1.5*.412, .5*.46) rectangle (2.5*.412, -3.5*.46);
            \draw (2.5*.412, -.5*.46) rectangle (3.5*.412, -3.5*.46);
        \end{tikzpicture}
\end{equation}
Note that the nonzero entries form rectangular blocks as shown. This is a result of the $\B_{T,i}$ being semi-infinite intervals extending either to the left or the right. Although the problem of computing the permanent of a general $(0,1)$-matrix is known to be \#P-complete \citep{valiant1979complexity}, \citet{christensen2023inference} proved that for matrices $A$ defined by \eqref{eq:matrixeq}, the permanent can be computed in polynomial time.

\section[Getting started with perms: a toy problem]{Getting started with \texttt{perms}: a toy problem}\label{sec:basicusage}
In this section we demonstrate how to use \texttt{perms} in practice. The basic functionality is illustrated with the following tractable toy problem. Let $n=100$, and let $0=t_1<t_2<\cdots<t_n=1$ be equally spaced points on the unit interval, and sample $X_1, \dots, X_n\sim\mathrm{Uniform}[0,1]$ independently. The task is to estimate $\log\prob(E)$, where $E$ is the event
$$E = \bigcap_{i=1}^{n/2}\{X_i > t_i\} \cap\bigcap_{i=n/2+1}^{n}\{X_i \leq t_i\}.$$
That is, the event that the first half of the $X_i$ are less than $t_i$, and that the second half is greater than $t_i$. In the notation of Section \ref{sec:intro}, the toy problem corresponds to estimating $\log \prob(Y=y)$, where $y=(0_{n/2}, 1_{n/2})^{\top}$, $X_i \overset{\text{ind}}{\sim} \text{Unif}[0,1]$, $T_i = t_i$, and there are no parameters $\theta$. Here, $0_{n/2}$ and $1_{n/2}$ denote row vectors of length $n/2$ containing only zeros and ones, respectively. 

Note that by independence, we can compute $\log\prob(E)$ directly as
$$\log\prob(E) = \sum_{i=1}^{n/2}\log(1 - t_i) + \sum_{i=n/2+1}^n\log t_i = -30.3751.$$
Hence, we see that $E$ is an extremely rare event whose probability cannot be estimated naively as a relative frequency of samples $X^{(s)}$ landing in the event, since this estimate would always yield zero in practice. However, using \texttt{perms}, we can employ the estimator \eqref{eq:logMLest}. As we shall see, a large proportion of simulated samples $X^{(s)}$ will satisfy $w(X^{(s)};\B) > 0$, and will therefore contribute to the estimate.

In the following two sections we will show how to solve the toy problem using \texttt{perms}. Since \texttt{perms} is implemented both as a Python library and an R package, we treat these separately, starting with Python. Readers only interested in the R implementation can skip to Section \ref{rinfosec}, as it is self-contained.

\subsection[Python]{Python}\label{basicpython}
The easiest way to install the Python version of \texttt{perms} is via the Python package index (\href{https://pypi.org/}{pypi.org}), which requires having \texttt{pip} installed. To install the newest version of \texttt{perms} using \texttt{pip}, we enter the following in the terminal:
\begin{CodeChunk}
\begin{CodeInput}
pip install perms
\end{CodeInput}
\end{CodeChunk}
Note that the Python version of \texttt{perms} is not available as a binary distribution, and hence \texttt{pip} requires a C compiler available on the target machine. Having installed \texttt{perms}, we import all relevant libraries:
\begin{CodeChunk}
\begin{CodeInput}
import numpy as np
import perms
\end{CodeInput}
\end{CodeChunk}
We now address the toy problem. We begin by specifying the data:
\begin{CodeChunk}
\begin{CodeInput}
n = 100
t = np.linspace(0, 1, n)
y = np.concatenate((np.zeros(n // 2), np.ones(n // 2))).astype('int32')
\end{CodeInput}
\end{CodeChunk}
Next, we generate \code{S = 20,000} random samples of $X$ and collect these in an \code{S} $\times$ \code{n} array \code{X_samples}:
\begin{CodeChunk}
\begin{CodeInput}
S = 20_000
X_samples = np.random.random((S, n))
\end{CodeInput}
\end{CodeChunk}
The log permutation numbers for each row of \code{X_samples} can be calculated with a single function call:
\begin{CodeChunk}
\begin{CodeInput}
log_perms = perms.get_log_perms(X_samples, t, y, False)
\end{CodeInput}
\end{CodeChunk}
Here, the final argument is a Boolean variable determining whether runtine information and debug messages will be printed to the terminal. The function \code{get_log_perms} returns the value \code{NaN} for any vanishing permutation number. That is, for any iteration with no contribution to the estimator \eqref{eq:logMLest}. We can therefore easily find how many nonzero permutation numbers were calculated as follows:
\begin{CodeChunk}
\begin{CodeInput}
num_nonzero_perms = np.sum(~np.isnan(log_perms))
\end{CodeInput}
\end{CodeChunk}
Having calculated \code{log_perms}, the resulting estimate of the log marginal likelihood can be computed with another function call:
\begin{CodeChunk}
\begin{CodeInput}
log_ML = perms.get_log_ML(log_perms, n, False) 
\end{CodeInput}
\end{CodeChunk}
Again, the final Boolean argument refers to the printing of runtime information and debug messages. The function \code{get_log_ML} calculates the estimate of the log marginal likelihood by applying the \code{LogSumExp} function to \code{log_perms} (ignoring any entries with value \code{NaN}), and subtracting $\log$\,\code{S} and $\log($\code{n}$!)$. This is because we divide by $S$ and $n!$ in \eqref{eq:logMLest}. 

The results of ten runs of the above code are given in Table~\ref{tab:toyproblempython}, along with the respective time they took to run. As we can see, we get a fast and accurate estimate in each run, with low variance. Also, note that all samples have nonzero permutation numbers in all runs. This is in stark contrast to the naive empirical estimate (which counts the frequency of samples landing in the event), in which we would expect every sample to get rejected.

\begin{table}[t!]
\centering
\begin{tabular}{cccc}
\hline
Run & log ML & proportion (\%) & time (s) \\
\hline
\phantom{1}1 & -30.3978 & 100 & 4.89 \\
\phantom{1}2 & -30.3938 & 100 & 5.00 \\
\phantom{1}3 & -30.4052 & 100 & 4.98 \\
\phantom{1}4 & -30.3844 & 100 & 4.70 \\
\phantom{1}5 & -30.3528 & 100 & 5.22 \\
\phantom{1}6 & -30.3833 & 100 & 5.31 \\
\phantom{1}7 & -30.3211 & 100 & 5.04 \\
\phantom{1}8 & -30.3768 & 100 & 5.20 \\
\phantom{1}9 & -30.4214 & 100 & 5.19 \\
10 & -30.3989 & 100 & 4.98 \\
 \hline
\end{tabular}
\caption{\label{tab:toyproblempython} Results from ten runs to solve the toy problem using Python, with \code{n = 100} and \code{S = 20,000}. Here, the log ML column contains the estimates of the log marginal likelihood, the proportion column contains the proportions of nonzero permutation numbers and the time column contains the run times in seconds.}
\end{table}

\subsection[R]{\texttt{R}}\label{rinfosec}
The easiest way to install the R version of \texttt{perms} is via the comprehensive R archive network (\href{https://cran.r-project.org}{cran.r-project.org}). The following code downloads and installs the latest version of \texttt{perms}: 
\begin{CodeChunk}
\begin{CodeInput}
R> install.packages("perms")
\end{CodeInput}
\end{CodeChunk}
Having installed \texttt{perms}, we now address the toy problem. We begin by specifying the data:
\begin{CodeChunk}
\begin{CodeInput}
R> library(perms)
R> n = 100
R> t = seq(0, 1, length.out = n)
R> y = c(rep(0, n / 2), rep(1, n / 2))
\end{CodeInput}
\end{CodeChunk}
Next, we generate \code{S = 20,000} random samples of $X$ and collect these in an \code{S} $\times$ \code{n} matrix \code{X_samples}:
\begin{CodeChunk}
\begin{CodeInput}
R> S = 20000
R> X_samples = matrix(runif(n * S), nrow = S, ncol = n)
\end{CodeInput}
\end{CodeChunk}
The log permutation numbers for each row of \code{X_samples} can be calculated with a single function call:
\begin{CodeChunk}
\begin{CodeInput}
R> log_perms = get_log_perms(X_samples, t, y, debug = FALSE, 
                            parallel = FALSE, num_cores = NULL)
\end{CodeInput}
\end{CodeChunk}
Here, the \code{debug} argument is a Boolean variable determining whether runtime information and debug messages will be printed to the terminal. As for the two last arguments, the code is executed in parallel whenever \code{parallel = TRUE}, and the number of cores is chosen automatically unless specified by \code{num_cores}. The function \code{get_log_perms} returns the value \code{NA} for any vanishing permutation number. That is, for any iteration with no contribution to the estimator \eqref{eq:logMLest}. We can therefore easily find how many nonzero permutation numbers were calculated as follows:
\begin{CodeChunk}
\begin{CodeInput}
R> num_nonzero_perms = sum(!is.na(log_perms))
\end{CodeInput}
\end{CodeChunk}

Having calculated \code{log_perms}, the resulting estimate of the log marginal likelihood can be computed with another function call:
\begin{CodeChunk}
\begin{CodeInput}
R> log_ML = get_log_ML(log_perms, n, debug = FALSE) 
\end{CodeInput}
\end{CodeChunk}
Again, the \code{debug} argument refers to the printing of debug messages. The function \code{get_log_ML} calculates the estimate of the log marginal likelihood by applying the \code{LogSumExp} function to \code{log_perms} (ignoring any entries with value \code{NA}), and subtracting $\log$\,\code{S} and $\log($\code{n}$!)$. This is because we divide by $S$ and $n!$ in \eqref{eq:logMLest}. 

The results of ten runs of the above code are given in Table~\ref{tab:toyproblemR}, along with the respective time they took to run. As we can see, we get a fast and accurate estimate in each run, with low variance. Note that the estimated log marginal likelihoods are not precisely the same as in Table~\ref{tab:toyproblempython} since the random number generators used by R and Python differ. Also, note that all samples have nonzero permutation numbers in all runs. This is in stark contrast to the naive empirical estimate (which counts the frequency of samples landing in the event), in which we would expect every sample to get rejected.

\begin{table}[t!]
\centering
\begin{tabular}{cccc}
\hline
Run & log ML & proportion (\%) & time (s) \\
\hline
\phantom{1}1 & -30.3604 & 100 & 5.12 \\
\phantom{1}2 & -30.4082 & 100 & 4.87 \\
\phantom{1}3 & -30.2962 & 100 & 5.09 \\
\phantom{1}4 & -30.3685 & 100 & 5.01 \\
\phantom{1}5 & -30.4039 & 100 & 5.02 \\
\phantom{1}6 & -30.4482 & 100 & 5.23 \\
\phantom{1}7 & -30.4157 & 100 & 5.34 \\
\phantom{1}8 & -30.3428 & 100 & 4.96 \\
\phantom{1}9 & -30.3859 & 100 & 5.30 \\
10 & -30.3833 & 100 & 5.09 \\
 \hline
\end{tabular}
\caption{\label{tab:toyproblemR} Results from ten runs to solve the toy problem using R, with \code{n = 100} and \code{S = 20,000}. Here, the log ML column contains the estimates of the log marginal likelihood, the proportion column contains the proportions of nonzero permutation numbers and the time column contains the run times in seconds.}
\end{table}

\section[Estimating marginal likelihoods with perms]{Estimating marginal likelihoods with \texttt{perms}}\label{sec:marginal_likelihoods_perms} We now show how \texttt{perms} is used to compute MLs for general problems with binary responses by studying two examples. Although \texttt{perms} is a likelihood free and thus highly general, we have deliberately selected examples for which alternative methods can be  used to estimate the marginal likelihood. These choices of examples enable us to assess the performance, ease of use and output of \texttt{perms} in comparison to established approaches. In Section \ref{BC}, we consider a standard Bayesian logistic regression model for binary classification. In Section \ref{bioassayml} we consider a bioassay problem with a Dirichlet process prior. In the interest of brevity, we only consider the Python version of \texttt{perms}.

\subsection[Bayesian logistic regression]{Bayesian logistic regression}\label{BC}
We now study the first of our two example problems, namely Bayesian logistic regression. Let $z_i\in\reals^p$ be the covariates associated to the $i$th sample (including a constant covariate for the intercept). In logistic regression, the model for the binary outcome $Y_i$ is
$$\prob(Y_i = 1\mid \theta) =  \sigma(\theta^\top z_i),$$
where $\sigma(a) = 1/(1 + \exp(-a))$ is the logistic sigmoid function. In the general framework of Section \ref{sec:intro}, this model corresponds to letting $T_i = \theta^\top z_i$, $X_i\sim\Logistic(1)$ and $Y_i = \one\{X_i \leq T_i\}$. To keep the analysis simple, we use independent standard normal priors for all the components of $\theta$, 
$$\theta \sim \N(0, I_p).$$
We now apply this model to the Iris data set \citep{anderson1936species}, analysed first by \citet{fisher1936use}. The data comprise 50 samples from three distinct Iris species, namely Iris setosa, Iris virginica, and Iris versicolor. For each sample, four characteristics were measured in centimeters: the length and width of both sepals and petals. Hence the total number of covariates, including the intercept, is $p=5$. The goal of our analysis is to classify whether a flower is of type setosa or not. All the data were rescaled to have mean zero and variance 1 before the analysis. Hence, the standard normal prior is somewhat informative.

To estimate the marginal likelihood using \texttt{perms}, we first generate an \code{S} $\times$ \code{p} matrix \code{theta} of standard normal random variables, each row of which is a prior sample. Recall that \code{S} is the number of samples used in the estimator. Next, let \code{Z} be the \code{n} $\times$ \code{p} matrix of covariates. We generate the \code{S} $\times$ \code{n} matrix of thresholds \code{T} by passing each row of \code{theta} through the observation model:
\begin{CodeChunk}
\begin{CodeInput}
T = theta @ Z.T
\end{CodeInput}
\end{CodeChunk}
Here, \code{@} refers to matrix multiplication. Finally, we create the \code{S} $\times$ \code{n} matrix of latent variables \code{X} by sampling from the standard logistic distribution,
\begin{CodeChunk}
\begin{CodeInput}
X_samples = np.random.logistic(size = (S, n))
\end{CodeInput}
\end{CodeChunk}
The log permutation numbers and the log ML can now be computed via two functions with \texttt{perms}:
\begin{CodeChunk}
\begin{CodeInput}
log_perms = perms.get_log_perms(X_samples, T, y, 0)
log_ML = perms.get_log_ML(log_perms, n, 0)
\end{CodeInput}
\end{CodeChunk}

Having seen how the ML can be estimated using \texttt{perms}, we will now consider the other two benchmark approaches. Since the model is tractable, we can employ conventional likelihood-based estimators of the ML. The first of these, which we shall refer to as the naive estimator, simply averages the likelihood function over the prior distribution,
$$\hat\pi_{\mathrm{naive},S'}(y) = \frac1{S'}\sum_{s=1}^{S'} \pi(y\mid\theta^{(s)}),$$
where $\theta^{(s)}\sim \pi(\theta)$ independently for $s=1, \dots, S'$. We remark that the naive estimator requires a tractable likelihood function, unlike \texttt{perms}. Although unbiased, conceptually simple and easy to use, the naive estimator is known to have large variance. A better alternative, given that the likelihood function is tractable, are bridge estimators \citep{meng1996simulating}, which require samples $\theta^{(s, 1)}\sim\pi(\theta)$ from the prior and $\theta^{(s,2)}\sim\pi(\theta\mid y)$ from the posterior. The general near-optimal bridge estimator takes the form
$$\hat\pi_{\mathrm{bridge}, S'}(y) = \frac{\sum_{s=1}^{S'}\pi(y\mid\theta^{(s, 1)})^{1/2}}{\sum_{s=1}^{S'}\pi(y\mid\theta^{(s,2)})^{-1/2}}.$$
This estimator is biased, but outperforms the naive estimator in most cases. Note that $\hat\pi_{\mathrm{bridge}, S'}(y)$ requires samples from the posterior, which we acquired via the Metropolis-Hastings algorithm with an isotropic Gaussian proposal,
$$\theta'\mid\theta \sim N(\theta, \sigma^2 I_p),$$
with $\sigma = 0.7$, found by trial and error. This proposal yielded an acceptance rate of approximately 23\%, which is close to the optimal rate \citep{gelman1997weak}. In all runs, an effective sample size of more than one thousand was achieved for each entry of $\theta$, with a 10\% burn-in rate.

\begin{figure}
    \centering
    \includegraphics{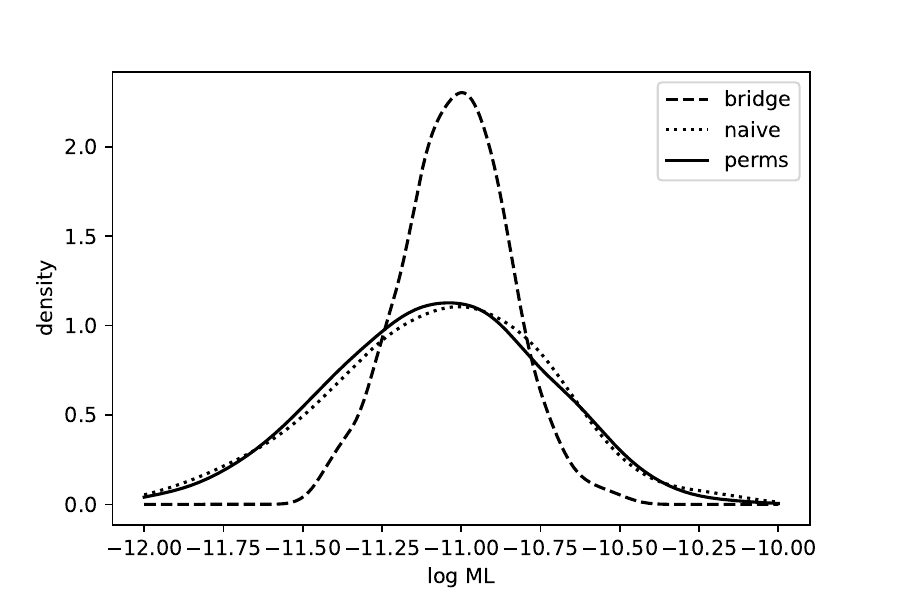}
    \caption{\label{fig:master_example} Kernel density estimates of the distribution of the log ML estimates using \texttt{perms} (solid curve), the naive estimator (dotted curve) and bridge estimator (dashed curve), with $M=250$ runs.}
\end{figure}

We now compare the performance of \texttt{perms}, the naive estimator and the bridge estimator over $M=250$ independent runs. For \texttt{perms}, we have used \code{S = 50,000} samples from the observation model in each run. A priori, it is not obvious how one should compare the performance of \texttt{perms} with that of the naive or the bridge estimators. Indeed, for the latter two, each sample contributes to the estimate, whereas for \texttt{perms}, a substantial proportion (approximately 57\% in our simulations) of the samples have zero permutation number and are thus immediately discarded. To ensure a fair comparison, in each of the $M=250$ runs, we set the number of samples in the naive and bridge estimators to be the number of non-zero permutation numbers obtained using \texttt{perms}. Figure \ref{fig:master_example} displays kernel density plots of the outputs of the three methods. As we can see, all three methods converge on the same value for the log ML. 

\begin{table}[t!]
\centering
\begin{tabular}{cccc}
\hline
\ & Average & Standard deviation & Avg. running time (s)\\ \hline
\texttt{perms} & -11.077 & 0.328 & 20.432\\
Naive estimator & -11.072 & 0.347 & 0.054\\
Bridge estimator & -11.018 & 0.167 & 0.180\\
\hline
\end{tabular}
\caption{\label{tab:masteraxample} The average value, standard deviation and average running time of the outputs from \texttt{perms}, the naive estimator, and the bridge estimator applied to the Iris data set. The values are computed from $M=250$ independent runs. }
\end{table}

Table \ref{tab:masteraxample} displays the average values and standard deviations of the outputs of the three competing methods, along with their average running times. We see that the estimation accuracy of \texttt{perms} slightly surpasses that of the naive estimator, though it falls short of the  accuracy achieved by the bridge estimator. As for the running time, the two competing methods are faster than \texttt{perms} by a significant factor. The discrepancy in performance between \texttt{perms} and the competitors can be attributed to the inherent generality of \texttt{perms}, which comes at a computational and statistical cost. Indeed, \texttt{perms} does not rely on the likelihood function to compute the marginal likelihood.

\subsection[Nonparametric bioassay]{Nonparametric bioassay}\label{bioassayml}
We now move on to our second example, namely a bioassay problem with a Dirichlet process prior. Such models were first studied by \citet{antoniak1974mixtures}. Although there exist many MCMC methods for posterior inference for this type of problem (see e.g. \citet{doss1994bayesian}), little work has gone into estimating the ML.

The data for the problem are given in Table~\ref{tab:bioassay_toy}. These data were generated by simulating $50$ independent data points from the mixture of Gaussians
$$\frac13\N(-2, 0.7^2) + \frac23\N(1,, 0.7^2)$$
for each of ten trials, and then counting how many of these were less than the given dosage level. Since there are $10$ trials, we have a total of $n=500$ observations. As prior, we use a Dirichlet process with standard normal base measure $P_0=\N(0,1)$ and concentration parameter $\alpha=1$, so $\proc = \Dir(1, \N(0,1))$. The is similar to that studied by \citet{christensen2023inference}, but here we consider a larger sample size of $n=500$, which is infeasible using \citeauthor{christensen2023inference}'s previous Python implementation.

\begin{table}[t!]
\centering
\begin{tabular}{ccc}
\hline
Dosage level & Number of successes & Number of trials\\ \hline
-1.00 & 10 & 50\\
-0.78 & 26 & 50\\
-0.56 & 10 & 50\\
-0.34 & 20 & 50\\
-0.11 & 20 & 50\\
\phantom{-}0.11 & 19 & 50\\
\phantom{-}0.33 & 29 & 50\\
\phantom{-}0.56 & 24 & 50\\
\phantom{-}0.78 & 31 & 50\\
\phantom{-}1.00 & 33 & 50\\ \hline
\end{tabular}
\caption{\label{tab:bioassay_toy} The dosage levels, number of successes and number of trials for the bioassay problem.}
\end{table}

Note that bioassay data of the type given in Table \ref{tab:bioassay_toy} is somewhat different than datasets for binary classification. Indeed, rather than listing each experiment $(t_i, y_i)$ individually, for $i=1, \dots, 500$, the data instead summarise how many trials were performed at each level, and how many of these resulted in successes. Hence, the functions \code{get_log_perms} and \code{get_log_ML} cannot be used immediately when the practitioner wishes to analyse bioassay data. In particular, the repeated trials introduce an additional factor of binomial coefficient to the ML (see the first remark on page 8 in \citet{christensen2023inference}). To save the user from having to reformat the data and account for these binomial coefficients themselves, we have instead equipped \texttt{perms} with an interface specifically designed for bioassay problems, which takes data of the form of Table \ref{tab:bioassay_toy} as input directly.

To compute the log ML using \texttt{perms}, we collect the numbers from Table~\ref{tab:bioassay_toy} into three \texttt{numpy} arrays called \code{levels}, \code{successes} and \code{trials}, respectively. We also define the number \code{num_trials} to be the number of different trials, that is, the number of rows in Table~\ref{tab:bioassay_toy}. To apply the estimator \eqref{eq:logMLest} using \texttt{perms}, we generate the \code{S} $\times$ \code{n} \texttt{numpy} array \code{X_samples}, whose rows are marginal Dirichlet process samples. Since our data are in the form of bioassay data, we calculate the log permutation numbers \code{log_perms} by calling the \code{get_log_perms_bioassay} function as follows:
\begin{CodeChunk}
\begin{CodeInput}
log_perms = perms.get_log_perms_bioassay(X_samples, levels, successes, 
                                                        trials, False) 
\end{CodeInput}
\end{CodeChunk}
Like \code{get_log_perms}, which we used for the previous two problems, this function also returns \code{NaN} for any vanishing permutation number. Having calculated \code{log_perms}, we can compute the resulting log marginal likelihood estimate \code{log_ML} with a single function call:
\begin{CodeChunk}
\begin{CodeInput}
log_ML = perms.get_log_ML_bioassay(log_perms, successes, trials, False)
\end{CodeInput}
\end{CodeChunk}
The function \code{get_log_ML_bioassay} works like \code{get_log_ML}, but since there are repeated trials (multiple trials at a single dosage level), \code{get_log_ML_bioassay} also adds the appropriate constants involving binomial factors, as remarked earlier.

For nonparametric models, there is no likelihood function available and thus neither the naive nor the bridge estimator can be employed. However, one natural candidate for estimating the ML is via stick-breaking approximations. Let $\alpha > 0$ and $P_0$ be a probability measure. \citet{sethuraman1994constructive} showed that if $B_1, B_2, \ldots \overset{\mathrm{iid}}{\sim}\mathrm{Beta}(1, \alpha)$, $\xi_1, \xi_2, \ldots \overset{\mathrm{iid}}{\sim} P_0$ and
$$w_i = B_i\prod_{j=1}^{i-1}(1 - B_j)$$
for all $i$, then
\begin{equation}\label{eq:stickbreaking}
    P = \sum_{i=1}^\infty w_i\delta_{\xi_i}\sim\mathrm{Dir}(\alpha, P_0)
\end{equation}
almost surely, where $\delta_\xi$ is the degenerate probability measure at the point $\xi$. Now, let $m$ denote the number of dosage levels and let $d$ denote the $m$-dimensional vector of dosage levels. Furthermore, let $\rho$ and $\tau$ denote the $m$-dimensional vectors of the number of successes and the number of trials at each level, respectively. The probability of observing the data $y$ given the (truncated) distribution $P$ is
\begin{equation}\label{eq:ML_stick}
    \pi(y\mid P) = \prod_{j=1}^m\left(\sum_{i:\xi_i \leq d_j}w_i\right)^{\rho_j}\left(\sum_{i:\xi_i > d_j}w_i\right)^{\tau_j - \rho_j}.
\end{equation}
By marginalising \eqref{eq:ML_stick} over $P$, we get an expression for the ML. In practice, however, we cannot include an infinite number of terms in \eqref{eq:stickbreaking}, but by truncating the sum in \eqref{eq:stickbreaking} at a big finite number $K$, we get an approximate expression $P_K$ for a single realisation of the Dirichlet process. In practice, one must choose $K$ large enough so that $\sum_{i=1}^K w_i \approx 1$. In our example, $K=1000$ consistently ensures this up to an error of $10^{-15}$. 
The ML can therefore be approximated using stick-breaking as
\begin{equation}
    \hat\pi_{\mathrm{stick}, S'}(y) = \frac1{S'}\sum_{s=1}^{S'}\pi(y\mid P_K^{(s)}),
\end{equation}
where $P^{(1)}, \ldots, P^{(S')}$ are independently generated $K$-truncated realisations of the Dirichlet process and each term in the sum is computed using \eqref{eq:ML_stick}. In practice, all terms are evaluated on a log scale and we use the \code{LogSumExp} function when computing the sums in \eqref{eq:ML_stick}.

\begin{figure}
    \centering
    \includegraphics{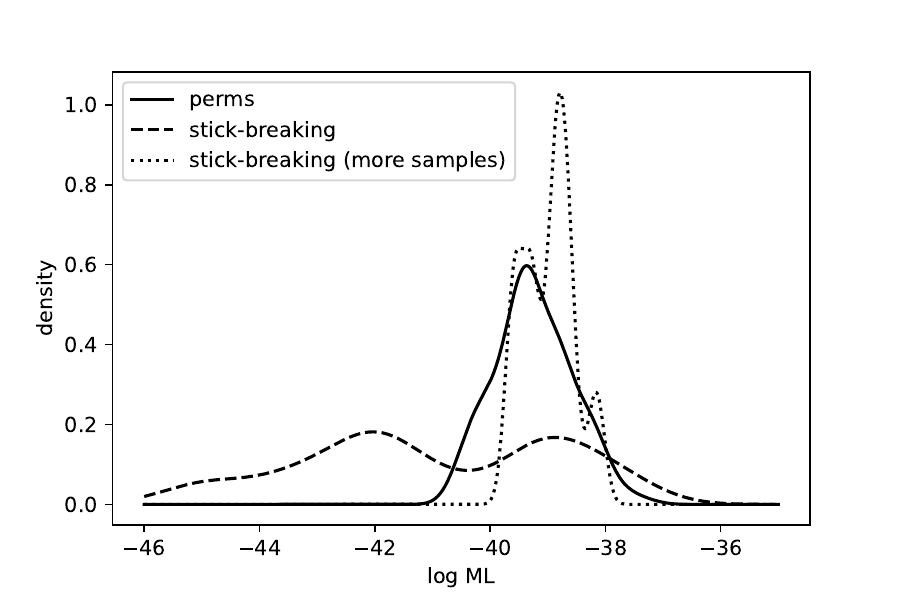}
    \caption{\label{fig:bioassay_fig} Kernel density estimates of the distribution of the log ML estimates using \texttt{perms} (solid curve), stick breaking (dotted curve), and stick breaking with reduced samples (dashed curve), with $M=250$ runs..}
\end{figure}

We estimated the ML using \texttt{perms} and stick-breaking with $S=20,000$ in $M=250$ separate runs. Since neither \texttt{perms} nor stick-breaking approximations have access to the likelihood function, they can be compared more fairly than the approaches in the previous section (i.e. the naive and bridge estimators). We therefore set $S' = S = 20,000$ for the number of stick-breaking approximations to compare the two methods with approximately the same computational budget. For completeness' sake, we also included a separate simulation where we set $S'$ to be the number of non-zero permutation numbers, so that both methods had the same number of contributing terms. In Figure \ref{fig:bioassay_fig}, we see kernel density estimates based on the results of the different runs.

The average values and standard deviations of the two methods are given in Table \ref{tab:permsMLbioassay}. Here we see that \texttt{perms} and stick-breaking yield similar performance given the same computational budget. Indeed, both the standard deviations and running times in the first two rows are relatively similar. That is, when comparing \texttt{perms} with another nonparametric estimator of the ML, we get similar results. Note, however, that \texttt{perms} does not require a stick-breaking representation, and is therefore even more general. Lastly, when $S'$ equals the number of non-zero permutation numbers, the accuracy of \texttt{perms} substantially outperforms that of stick-breaking, although naturally at a cost of increased running time.

\begin{table}[t!]
\centering
\begin{tabular}{cccc}
\hline
\ & Average & Standard deviation & Avg. running time (s)\\ \hline
\texttt{perms} & -39.263 & 0.634 & 3.538\\
Stick-breaking & -39.109 & 0.486 & 3.468\\
Stick-breaking (fewer samples) & -40.633 & 1.915 & 0.110\\
\hline
\end{tabular}
\caption{\label{tab:permsMLbioassay} The average values, standard deviations and average running times of the outputs from \texttt{perms} and the stick-breaking approximation in the nonparametric bioassay problem. The values are computed from $M=250$ independent runs.}
\end{table}

\section{Computation of permutation numbers}\label{sec:computation}
We now provide an overview of how \texttt{perms} computes $\per\,A$, where $A$ is defined as in \eqref{eq:matrixeq}. This is needed for understanding the implementation details covered in Section~\ref{sec:implementation_details}, which will explain the computational efficiency of \texttt{perms}. 

\subsection{Reduction}\label{sec:reduction31}
The key idea for computing $\per\,A$ is to \emph{reduce} the matrix $A$ to a smaller matrix $A'$ (with fewer rows) via of series of moves. The details of the reduction moves are given in \citet[supplementary material]{christensen2023inference}, but the main idea is that under certain conditions, we may remove rows of the matrix $A$ and merge rectangular blocks together. Since the reduction introduces rectangular matrices, we also need the definition of the permanent of an $m\times n$ matrix $A$ with $m\leq n$. This is given by
\begin{equation}\label{eq:permrect}
    \per\,A = \sum_{\sigma\in S_{n,m}}\prod_{i=1}^m a_{i,\sigma(i)},
\end{equation}
where $S_{n,m}$ is the set of $m$-permutations of the set $\{1, \dots, n\}$.

\begin{figure}
\scalebox{0.89}{
    \begin{tabular}{cccccc}
        &
        \begin{tikzpicture}[baseline={(0,0)}]
            \matrix[matrix of math nodes, left delimiter=(, right delimiter=), ampersand replacement=\&]{ 
                1 \& 1 \& 1 \& 1 \& 0 \& 0 \& 0 \\
                1 \& 1 \& 1 \& 1 \& 1 \& 0 \& 0 \\
                1 \& 1 \& 1 \& 1 \& 1 \& 0 \& 0 \\
                1 \& 1 \& 1 \& 1 \& 1 \& 1 \& 0 \\
                0 \& 1 \& 1 \& 1 \& 1 \& 1 \& 1 \\
                0 \& 0 \& 0 \& 0 \& 1 \& 1 \& 1 \\
                0 \& 0 \& 0 \& 0 \& 0 \& 1 \& 1 \\
                };
            %\draw (0,0) rectangle (2*.45, .51);
            \draw (-3.5*.45, 3.5*.51) rectangle (-2.5*.45,-.5*.51);
            \draw (-2.5*.45, 3.5*.51) rectangle (.5*.45, -1.5*.51);
            \draw (.5*.45, 2.5*.51) rectangle (1.5*.45, -2.5*.51);
            \draw (1.5*.45, .5*.51) rectangle (2.5*.45, -3.5*.51);
            \draw (2.5*.45, -.5*.51) rectangle (3.5*.45, -3.5*.51);
        \end{tikzpicture}
        &
        $\rightarrow$
        &
        \begin{tikzpicture}[baseline={(0,0)}]
            \matrix[matrix of math nodes, left delimiter=(, right delimiter=), ampersand replacement=\&]{ 
                1 \& 1 \& 1 \& 1 \& 0 \& 0 \& 0 \\
                1 \& 1 \& 1 \& 1 \& 1 \& 0 \& 0 \\
                1 \& 1 \& 1 \& 1 \& 1 \& 0 \& 0 \\
                1 \& 1 \& 1 \& 1 \& 1 \& 1 \& 0 \\
                0 \& 0 \& 0 \& 0 \& 1 \& 1 \& 1 \\
                0 \& 0 \& 0 \& 0 \& 0 \& 1 \& 1 \\
                };
            %\draw[densely dotted] (-3.5*.41, 2.5*.46) -- (3.5*.41, 2.5*.46);
            \draw (-3.5*.45, 3*.51) rectangle (-2.5*.45,-1*.51);
            \draw (-2.5*.45, 3*.51) rectangle (.5*.45, -1*.51);
            \draw (.5*.45, 2*.51) rectangle (1.5*.45, -2*.51);
            \draw (1.5*.45, 0*.51) rectangle (2.5*.45, -3*.51);
            \draw (2.5*.45, -1*.51) rectangle (3.5*.45, -3*.51);
        \end{tikzpicture}
        &
        $\rightarrow$
        &
        \begin{tikzpicture}[baseline={(0,0)}]
            \matrix[matrix of math nodes, left delimiter=(, right delimiter=), ampersand replacement=\&]{ 
                1 \& 1 \& 1 \& 1 \& 0 \& 0 \& 0 \\
                1 \& 1 \& 1 \& 1 \& 1 \& 0 \& 0 \\
                1 \& 1 \& 1 \& 1 \& 1 \& 0 \& 0 \\
                1 \& 1 \& 1 \& 1 \& 1 \& 1 \& 0 \\
                0 \& 0 \& 0 \& 0 \& 1 \& 1 \& 1 \\
                0 \& 0 \& 0 \& 0 \& 0 \& 1 \& 1 \\
                };
            %\draw[densely dotted] (-3.5*.41, 2.5*.46) -- (3.5*.41, 2.5*.46);
            \draw (-3.5*.45, 3*.51) rectangle (.5*.45, -1*.51);
            \draw (.5*.45, 2*.51) rectangle (1.5*.45, -2*.51);
            \draw (1.5*.45, 0*.51) rectangle (2.5*.45, -3*.51);
            \draw (2.5*.45, -1*.51) rectangle (3.5*.45, -3*.51);
        \end{tikzpicture} \vspace{3ex} \\
        $\rightarrow$
        &
        \begin{tikzpicture}[baseline={(0,0)}]
            \matrix[matrix of math nodes, left delimiter=(, right delimiter=), ampersand replacement=\&]{ 
                1 \& 1 \& 1 \& 1 \& 0 \& 0 \& 0 \\
                1 \& 1 \& 1 \& 1 \& 1 \& 0 \& 0 \\
                1 \& 1 \& 1 \& 1 \& 1 \& 0 \& 0 \\
                0 \& 0 \& 0 \& 0 \& 1 \& 1 \& 1 \\
                0 \& 0 \& 0 \& 0 \& 0 \& 1 \& 1 \\
                };
            %\draw[densely dotted] (-3.5*.41, 2.5*.46) -- (3.5*.41, 2.5*.46);
            \draw (-3.5*.45, 2.5*.51) rectangle (.5*.45, -.5*.51);
            \draw (.5*.45, 1.5*.51) rectangle (1.5*.45, -1.5*.51);
            \draw (1.5*.45, -.5*.51) rectangle (2.5*.45, -2.5*.51);
            \draw (2.5*.45, -.5*.51) rectangle (3.5*.45, -2.5*.51);
        \end{tikzpicture}
        &
        $\rightarrow$
        &
        \begin{tikzpicture}[baseline={(0,0)}]
            \matrix[matrix of math nodes, left delimiter=(, right delimiter=), ampersand replacement=\&]{ 
                1 \& 1 \& 1 \& 1 \& 0 \& 0 \& 0 \\
                1 \& 1 \& 1 \& 1 \& 1 \& 0 \& 0 \\
                1 \& 1 \& 1 \& 1 \& 1 \& 0 \& 0 \\
                0 \& 0 \& 0 \& 0 \& 1 \& 1 \& 1 \\
                0 \& 0 \& 0 \& 0 \& 0 \& 1 \& 1 \\
                };
            %\draw[densely dotted] (-3.5*.41, 2.5*.46) -- (3.5*.41, 2.5*.46);
            \draw (-3.5*.45, 2.5*.51) rectangle (.5*.45, -.5*.51);
            \draw (.5*.45, 1.5*.51) rectangle (1.5*.45, -1.5*.51);
            \draw (1.5*.45, -.5*.51) rectangle (3.5*.45, -2.5*.51);
        \end{tikzpicture}
        &
        $\rightarrow$
        &
        \begin{tikzpicture}[baseline={(0,0)}]
            \matrix[matrix of math nodes, left delimiter=(, right delimiter=), ampersand replacement=\&]{
                1 \& 1 \& 1 \& 1 \& 1 \& 0 \& 0 \\
                1 \& 1 \& 1 \& 1 \& 1 \& 0 \& 0 \\
                0 \& 0 \& 0 \& 0 \& 1 \& 1 \& 1 \\
                0 \& 0 \& 0 \& 0 \& 0 \& 1 \& 1 \\
                };
            %\draw[densely dotted] (-3.5*.41, 2.5*.46) -- (3.5*.41, 2.5*.46);
            \draw (-3.5*.45, 2*.51) rectangle (.5*.45, 0*.51);
            \draw (.5*.45, 2*.51) rectangle (1.5*.45, -1*.51);
            \draw (1.5*.45, 0*.51) rectangle (3.5*.45, -2*.51);
        \end{tikzpicture} \vspace{3ex} \\
        $\rightarrow$
        &
        \begin{tikzpicture}[baseline={(0,0)}]
            \matrix[matrix of math nodes, left delimiter=(, right delimiter=), ampersand replacement=\&]{
                1 \& 1 \& 1 \& 1 \& 1 \& 0 \& 0 \\
                1 \& 1 \& 1 \& 1 \& 1 \& 0 \& 0 \\
                0 \& 0 \& 0 \& 0 \& 0 \& 1 \& 1 \\
                };
            %\draw[densely dotted] (-3.5*.41, 2.5*.46) -- (3.5*.41, 2.5*.46);
            \draw (-3.5*.45, 1.5*.51) rectangle (.5*.45, -.5*.51);
            \draw (.5*.45, 1.5*.51) rectangle (1.5*.45, -.5*.51);
            \draw (1.5*.45, -.5*.51) rectangle (3.5*.45, -1.5*.51);
        \end{tikzpicture}
        &
        $\rightarrow$
        &
        \begin{tikzpicture}[baseline={(0,0)}]
            \matrix[matrix of math nodes, left delimiter=(, right delimiter=), ampersand replacement=\&]{
                1 \& 1 \& 1 \& 1 \& 1 \& 0 \& 0 \\
                1 \& 1 \& 1 \& 1 \& 1 \& 0 \& 0 \\
                0 \& 0 \& 0 \& 0 \& 0 \& 1 \& 1 \\
                };
            %\draw[densely dotted] (-3.5*.41, 2.5*.46) -- (3.5*.41, 2.5*.46);
            \draw (-3.5*.45, 1.5*.51) rectangle (1.5*.45, -.5*.51);
            \draw (1.5*.45, -.5*.51) rectangle (3.5*.45, -1.5*.51);
        \end{tikzpicture}
        &
        $\rightarrow$
        &
        \begin{tikzpicture}[baseline={(0,0)}]
            \matrix[matrix of math nodes, left delimiter=(, right delimiter=), ampersand replacement=\&]{
                1 \& 1 \& 1 \& 1 \& 1 \& 0 \& 0 \\
                1 \& 1 \& 1 \& 1 \& 1 \& 0 \& 0 \\
                };
            %\draw[densely dotted] (-3.5*.41, 2.5*.46) -- (3.5*.41, 2.5*.46);
            \draw (-3.5*.45, 1*.51) rectangle (1.5*.45, -1*.51);
        \end{tikzpicture}
    \end{tabular}}
    \caption{\label{fig:reduction} Reduction of the example matrix $A_1$ from \eqref{eq:favourite}.}
\end{figure}

In Figure~\ref{fig:reduction}, we see the reduction of the matrix $A_1$ from \eqref{eq:favourite}. The reduction ends with the matrix $A_1'$, given by
\begin{equation}\label{eq:favouritereducedfully}
A_1' = \begin{tikzpicture}[baseline={(0,0)}]
            \matrix[matrix of math nodes, left delimiter=(, right delimiter=), ampersand replacement=\&]{ 
                1 \& 1 \& 1 \& 1 \& 1 \& 0 \& 0 \\
                1 \& 1 \& 1 \& 1 \& 1 \& 0 \& 0 \\
            };
            \draw (-3.5*.412, 1*.46) rectangle (1.5*.412, -1*.46);
        \end{tikzpicture},
\end{equation}
which has a much simpler structure than the original matrix $A_1$. In fact, we can see straight away that $\per\,A_1' = 5\times 4 = 20$. 

Having computed $\per\, A'$ for the reduced matrix, the goal is to reverse the reduction to obtain the permanent of the original matrix $A$. Unfortunately, there is no direct recurrence relation which relates the permanent of $A'$ to that of the matrix which came before it in the reduction. However, this will be possible once we introduce the notion of \emph{subpermanents}.

\subsection{Subpermanents}\label{subpermsec}
\begin{figure}
    \begin{subfigure}{28ex}
        \centering
        \begin{tikzpicture}[baseline={(0,0)}]
            \matrix[matrix of math nodes, left delimiter=(, right delimiter=)]{ 
                1 & 1 & 1 & 1 & 0 & 0 & 0 \\
                1 & 1 & 1 & 1 & 1 & 0 & 0 \\
                1 & 1 & 1 & 1 & 1 & 0 & 0 \\
                1 & 1 & 1 & 1 & 1 & 1 & 0 \\
                0 & 1 & 1 & 1 & 1 & 1 & 1 \\
                0 & 0 & 0 & 0 & 1 & 1 & 1 \\
                0 & 0 & 0 & 0 & 0 & 1 & 1 \\
                };
            \draw (-3.5*.45, 3.5*.51) rectangle (-2.5*.45,-.5*.51);
            \draw (-2.5*.45, 3.5*.51) rectangle (.5*.45, -1.5*.51);
            \draw (.5*.45, 2.5*.51) rectangle (1.5*.45, -2.5*.51);
            \draw (1.5*.45, .5*.51) rectangle (2.5*.45, -3.5*.51);
            \draw (2.5*.45, -.5*.51) rectangle (3.5*.45, -3.5*.51);
            \draw[color=red](-3*.45,3*.51) circle (.19);
            \draw[color=red](-2*.45,2*.51) circle (.19);
            \draw[color=red](-1*.45,1*.51) circle (.19);
            \draw[color=red](0*.45,0*.51) circle (.19);
            \draw[color=red](1*.45,-1*.51) circle (.19);
            \draw[color=red](2*.45,-2*.51) circle (.19);
            \draw[color=red](3*.45,-3*.51) circle (.19);
        \end{tikzpicture}
        \caption{\label{fig:subperm_a}}
    \end{subfigure}
    \begin{subfigure}{28ex}
        \centering
        \begin{tikzpicture}[baseline={(0,0)}]
            \matrix[matrix of math nodes, left delimiter=(, right delimiter=)]{ 
                1 & 1 & 1 & 1 & 0 & 0 & 0 \\
                1 & 1 & 1 & 1 & 1 & 0 & 0 \\
                1 & 1 & 1 & 1 & 1 & 0 & 0 \\
                1 & 1 & 1 & 1 & 1 & 1 & 0 \\
                0 & 1 & 1 & 1 & 1 & 1 & 1 \\
                0 & 0 & 0 & 0 & 1 & 1 & 1 \\
                0 & 0 & 0 & 0 & 0 & 1 & 1 \\
                };
            \draw (-3.5*.45, 3.5*.51) rectangle (-2.5*.45,-.5*.51);
            \draw (-2.5*.45, 3.5*.51) rectangle (.5*.45, -1.5*.51);
            \draw (.5*.45, 2.5*.51) rectangle (1.5*.45, -2.5*.51);
            \draw (1.5*.45, .5*.51) rectangle (2.5*.45, -3.5*.51);
            \draw (2.5*.45, -.5*.51) rectangle (3.5*.45, -3.5*.51);
            \draw[color=red](-3*.45,1*.51) circle (.19);
            \draw[color=red](-2*.45,3*.51) circle (.19);
            \draw[color=red](-1*.45,2*.51) circle (.19);
            \draw[color=red](0*.45,0*.51) circle (.19);
            \draw[color=red](1*.45,-2*.51) circle (.19);
            \draw[color=red](2*.45,-1*.51) circle (.19);
            \draw[color=red](3*.45,-3*.51) circle (.19);
        \end{tikzpicture}
        \caption{\label{fig:subperm_b}}
    \end{subfigure}
    \begin{subfigure}{28ex}
        \centering
        \begin{tikzpicture}[baseline={(0,0)}]
            \matrix[matrix of math nodes, left delimiter=(, right delimiter=)]{ 
                1 & 1 & 1 & 1 & 0 & 0 & 0 \\
                1 & 1 & 1 & 1 & 1 & 0 & 0 \\
                1 & 1 & 1 & 1 & 1 & 0 & 0 \\
                1 & 1 & 1 & 1 & 1 & 1 & 0 \\
                0 & 1 & 1 & 1 & 1 & 1 & 1 \\
                0 & 0 & 0 & 0 & 1 & 1 & 1 \\
                0 & 0 & 0 & 0 & 0 & 1 & 1 \\
                };
            \draw (-3.5*.45, 3.5*.51) rectangle (-2.5*.45,-.5*.51);
            \draw (-2.5*.45, 3.5*.51) rectangle (.5*.45, -1.5*.51);
            \draw (.5*.45, 2.5*.51) rectangle (1.5*.45, -2.5*.51);
            \draw (1.5*.45, .5*.51) rectangle (2.5*.45, -3.5*.51);
            \draw (2.5*.45, -.5*.51) rectangle (3.5*.45, -3.5*.51);
            \draw[color=red](-3*.45,2*.51) circle (.19);
            \draw[color=red](-2*.45,0*.51) circle (.19);
            \draw[color=red](-1*.45,3*.51) circle (.19);
            \draw[color=red](0*.45,1*.51) circle (.19);
            \draw[color=red](1*.45,-2*.51) circle (.19);
            \draw[color=red](2*.45,-3*.51) circle (.19);
            \draw[color=red](3*.45,-1*.51) circle (.19);
        \end{tikzpicture}
        \caption{\label{fig:subperm_c}}
    \end{subfigure}
    
    \begin{subfigure}{28ex}
        \centering
        \begin{tikzpicture}[baseline={(0,0)}]
            \matrix[matrix of math nodes, left delimiter=(, right delimiter=)]{ 
                1 & 1 & 1 & 1 & 0 & 0 & 0 \\
                1 & 1 & 1 & 1 & 1 & 0 & 0 \\
                1 & 1 & 1 & 1 & 1 & 0 & 0 \\
                0 & 0 & 0 & 0 & 1 & 1 & 1 \\
                0 & 0 & 0 & 0 & 0 & 1 & 1 \\
                };
            \draw (-3.5*.45, 2.5*.51) rectangle (.5*.45,-.5*.51);
            \draw (.5*.45, 1.5*.51) rectangle (1.5*.45, -1.5*.51);
            \draw (1.5*.45, -.5*.51) rectangle (3.5*.45, -2.5*.51);
            \draw[color=red](-2*.45,2*.51) circle (.19);
            \draw[color=red](-1*.45,1*.51) circle (.19);
            \draw[color=red](0*.45,0*.51) circle (.19);
            \draw[color=red](1*.45,-1*.51) circle (.19);
            \draw[color=red](2*.45,-2*.51) circle (.19);
        \end{tikzpicture}
        \caption{\label{fig:subperm_d}}
    \end{subfigure}
    \begin{subfigure}{28ex}
        \centering
        \begin{tikzpicture}[baseline={(0,0)}]
            \matrix[matrix of math nodes, left delimiter=(, right delimiter=)]{ 
                1 & 1 & 1 & 1 & 0 & 0 & 0 \\
                1 & 1 & 1 & 1 & 1 & 0 & 0 \\
                1 & 1 & 1 & 1 & 1 & 0 & 0 \\
                0 & 0 & 0 & 0 & 1 & 1 & 1 \\
                0 & 0 & 0 & 0 & 0 & 1 & 1 \\
                };
            \draw (-3.5*.45, 2.5*.51) rectangle (.5*.45,-.5*.51);
            \draw (.5*.45, 1.5*.51) rectangle (1.5*.45, -1.5*.51);
            \draw (1.5*.45, -.5*.51) rectangle (3.5*.45, -2.5*.51);
            \draw[color=red](-3*.45,0*.51) circle (.19);
            \draw[color=red](-2*.45,2*.51) circle (.19);
            \draw[color=red](-1*.45,1*.51) circle (.19);
            \draw[color=red](1*.45,-1*.51) circle (.19);
            \draw[color=red](3*.45,-2*.51) circle (.19);
        \end{tikzpicture}
        \caption{\label{fig:subperm_e}}
    \end{subfigure}
    \begin{subfigure}{28ex}
        \centering
        \begin{tikzpicture}[baseline={(0,0)}]
            \matrix[matrix of math nodes, left delimiter=(, right delimiter=)]{ 
                1 & 1 & 1 & 1 & 0 & 0 & 0 \\
                1 & 1 & 1 & 1 & 1 & 0 & 0 \\
                1 & 1 & 1 & 1 & 1 & 0 & 0 \\
                0 & 0 & 0 & 0 & 1 & 1 & 1 \\
                0 & 0 & 0 & 0 & 0 & 1 & 1 \\
                };
            \draw (-3.5*.45, 2.5*.51) rectangle (.5*.45,-.5*.51);
            \draw (.5*.45, 1.5*.51) rectangle (1.5*.45, -1.5*.51);
            \draw (1.5*.45, -.5*.51) rectangle (3.5*.45, -2.5*.51);
            \draw[color=red](-3*.45,0*.51) circle (.19);
            \draw[color=red](0*.45,2*.51) circle (.19);
            \draw[color=red](1*.45,1*.51) circle (.19);
            \draw[color=red](2*.45,-2*.51) circle (.19);
            \draw[color=red](3*.45,-1*.51) circle (.19);
        \end{tikzpicture}
        \caption{\label{fig:subperm_f}}
    \end{subfigure}
    \caption{\label{fig:wpAexample} Top row: three elements of $\wp(A_1)$ for the matrix $A_1$ given in \eqref{eq:favourite}. Bottom row: three elements of $\wp(A_2)$ for the matrix $A_2$ given in \eqref{eq:favouritereduced}. The red circles denote chosen entries.}
\end{figure}
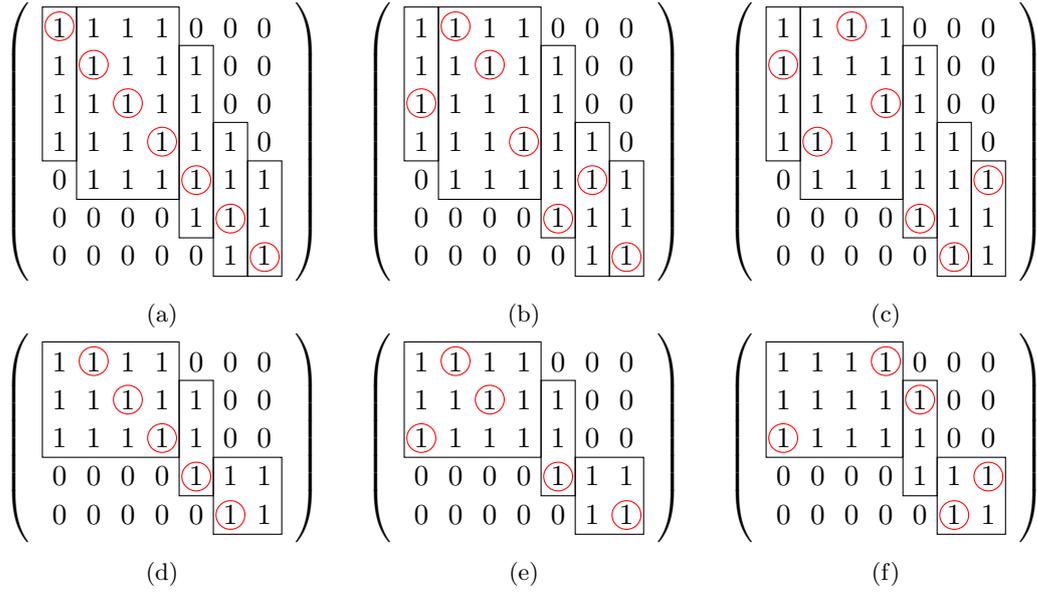

From the definition \eqref{eq:permrect}, the permanent of a matrix $A$ equals the cardinality of the set $\wp(A)$, where we define
\begin{equation}
    \wp(A) = \{\sigma \in S_{n,m} : a_{1,\sigma(1)} = \dots = a_{m,\sigma(m)} = 1\}.
\end{equation}
The elements of $\wp(A)$ have a useful geometric interpretation. They correspond to ways in which we can choose exactly $m$ entries $a_{ij}$ from the matrix $A$ such that all $a_{ij}=1$ and all the rows provide precisely one entry. In Figure~\ref{fig:wpAexample}, we see a visual representation of three elements of $\wp(A_1)$ for the matrix $A_1$ from \eqref{eq:favourite}. We also see three elements of $\wp(A_2)$ for the matrix $A_2$, given by
\begin{equation}\label{eq:favouritereduced}
    A_2 = \begin{tikzpicture}[baseline={(0,0)}]
            \matrix[matrix of math nodes, left delimiter=(, right delimiter=)]{ 
                1 & 1 & 1 & 1 & 0 & 0 & 0 \\
                1 & 1 & 1 & 1 & 1 & 0 & 0 \\
                1 & 1 & 1 & 1 & 1 & 0 & 0 \\
                0 & 0 & 0 & 0 & 1 & 1 & 1 \\
                0 & 0 & 0 & 0 & 0 & 1 & 1 \\
                };
            \draw (-3.5*.412, 2.5*.46) rectangle (.5*.412,-.5*.46);
            \draw (.5*.412, 1.5*.46) rectangle (1.5*.412, -1.5*.46);
            \draw (1.5*.412, -.5*.46) rectangle (3.5*.412, -2.5*.46);
        \end{tikzpicture},
\end{equation}
which appears in the reduction in Figure~\ref{fig:reduction}.

By considering how many chosen entries of an element $\sigma\in\wp(A)$ are in the left most and right most blocks, we obtain the subpermanents. More precisely, define the sets $\wp_{r,s}(A)$ by
\begin{equation}\label{eq:wprsA}
    \wp_{r,s}(A) = \left\{\sigma \in \wp(A) : \begin{tabular}{l}$r$ entries from left most block \\ $s$ entries from right most block\end{tabular}\right\}.
\end{equation}
We define the $(r,s)$ \emph{subpermanent} $\per_{r,s}(A)$ as the cardinality of the set $\wp_{r,s}(A)$. That is,
\begin{equation*}
    \per_{r,s}(A) = \#\wp_{r,s}(A).
\end{equation*}
Note that the permanent is the sum of all subpermanents,
\begin{equation}\label{eq:subpermssum}
    \per\, A = \sum_{r,s}\per_{r,s}(A).
\end{equation}

From the definition \eqref{eq:wprsA}, we see that the permutations in Figure~\ref{fig:subperm_d} and \ref{fig:subperm_e} represent elements in $\wp_{3, 1}(A_2)$, whilst the permutation in Figure~\ref{fig:subperm_f} represents an element in $\wp_{2,2}(A_2)$.

As a simple example, let us compute the subpermanents of the matrix $A_1'$ from \eqref{eq:favouritereducedfully}. Since the left most block is the only block, the right most block has height zero implicitly. We are thus forced to select two entries from the left most block. Hence we see that $\per_{2,0}(A_1') = 20$ and $\per_{r,s}(A_1') = 0$ for all other choices of $r$ and $s$.

Having computed all subpermanents of the reduced matrix $A'$, we can use the recurrence relations given in \citet[supplementary material]{christensen2023inference} to derive the subpermanents of the matrix in the previous step of the reduction. Repeating this process, we reverse the reduction fully and end up with the subpermanents $\per_{r,s}(A)$ of the original matrix $A$. By \eqref{eq:subpermssum}, we recover $\per\, A$ by summing all subpermanents $\per_{r,s}(A)$ of $A$.

We summarise the key steps of the algorithm for computing $\per\,A$ as follows:
\begin{enumerate}
    \item Using the moves given in \citet[supplementary material]{christensen2023inference}, reduce the matrix $A$ to the reduced matrix $A'$.

    \item Calculate the subpermanents $\per_{r,s}(A')$ of the reduced matrix $A'$.

    \item Use the recurrence relations from \citet[supplementary material]{christensen2023inference} to reverse the reduction, yielding the subpermanents $\per_{r,s}(A)$ in the final step.

    \item Return $\per\,A = \sum_{r,s}\per_{r,s}(A)$.
\end{enumerate}

We recall that this algorithm computes the permutation number of a single random sample $X^{(s)}$ in the estimator \eqref{eq:logMLest}.

\section{Implementation details}\label{sec:implementation_details}
In this section we give details about the inner workings of \texttt{perms}. The vast part of the code is centred around computing permanents of block rectangular matrices, which is done using the algorithm given in Section~\ref{sec:computation}. In what follows, we explain the various optimisation steps taken for these computations. As the functions \code{get_log_perms} and \code{get_log_perms_bioassay} only differ in the representation of their inputs, we consider only the function \code{get_log_perms}.

%Here \code{n} denotes the sample size of the data, and \code{S} denotes the number of samples drawn from the data model. The \code{S} $\times$ \code{n} matrix \code{X} should contain a sample from the data model in each row. The vector or matrix valued input \code{t} represents the ?? and y represents ?? [xx legg til info her xx]. If \code{debug} is \code{True}, run time debug information is printed to the terminal. 
\subsection{Programming language}
Python and R are arguably the most used programming languages for statistical computing. Nevertheless, when confronted with computationally demanding tasks, these languages tend to be considerably slower than languages like C \citep[see][for a comparison between Python and C++, the latter language being an extension of C]{pythonversusc}. 

The divergence in speed between C on one hand and Python and R on the other is underpinned by several factors. Firstly, Python and R are interpreted languages, meaning that the human readable Python or R code is parsed and converted into CPU instructions in real-time. In comparison, C is a compiled language, enabling a range of optimisation procedures to be executed during compilation. Secondly, C is strongly typed, whilst Python and R are weakly typed, incurring overhead during execution due to type checking. Thirdly, C is a lower level language, providing the programmer with more direct access to hardware resources and system level operations. This level of control allows for fine-tuning and optimisation which is not as feasible in higher level languages like Python or R. For these reasons, we have decided to write \texttt{perms} in C. To make the C code available from Python and R, we have used the respective C interfaces of the two languages \citep[see][respectively]{noauthor_pythonc_2023,the_r_core_team_writing_2023}.

\subsection{Representation of data}
Recall that the function \code{get_log_perms} takes arguments \code{X}, \code{t}, \code{y}, \code{n}, \code{S} and \code{debug}. The function iterates through each row of \code{X}, i.e., each sample $X^{(s)}$ from the data model. At each iteration, the matrix $A$ as in \eqref{eq:matrixeq} is created. From Definition 3.2 and Proposition 3.1 in \citet{christensen2023inference}, we can parametrise the matrix $A$ compactly by the number of rows $m$, the number of blocks $k$, and the three vectors $\alpha\in\Z_{>0}^k, \beta, \gamma\in\Z_{\geq 0}^{k-1}$, which determine the dimensions of the blocks. As $k\leq n$, this representation of $A$ reduces memory use from $\mathcal{O}(n^2)$ to $\mathcal{O}(n)$. The number of rows $m$ is a priori equal to the sample size $n$, but will change during the actual computation of the permanent of $A$ due to the reduction steps. In the code, the matrix $A$ is generated via the function \code{get_alpha_beta_gamma}, which computes \code{alpha}, \code{beta}, \code{gamma}, \code{m} and \code{k}. 
%MER OM AT ALT LAGRES I EN SINGLE ARRAY OSV OG AT VI BARE OPPDATERER VERDIENE DENS. GJERNE OGSÅ OM AT ALT HAR FIXED LENGTH OG AT VI BRUKER K SOM BOUND FOR ANTALL RELEVANT ENTRIES OSV. -> tas senere

%\subsection{Iterative procedure for sub-permanents}
%Once the parametrisation of the current block rectangular matrix is in place, the function proceeds to compute the permanent of the matrix. This is done following the algorithm of <dennispaper>, by iteratively applying a set of instructions to reduce the size of the matrix until its sub-permanents are easily computable. 
%
%<Dennis-forklaring og eksempel> 
%
%Among the reduction steps are \code{top_trim},..., etc. While the current block rectangular matrix is reduced, the history of the applied reductions are stored in a vector indicating which operation was performed.
%
%Once the current block rectangular matrix has been fully reduced, the sub-permanents of the resulting matrix is computed by the function \code{sparse_get_reduced_subperms}. Then the process of reduction is reversed, in which the permanent of the (larger) matrix before a reduction step is computed using the sub-permanents of the matrix after the reduction step. 
%

\subsection{Discarding vanishing permanents}\label{ref:discarding}
In most real applications, the prior process $\proc$ is agnostic regarding the observed data, and so the estimator \eqref{eq:logMLest} will have a substantial proportion of samples $X^{(s)}$ with $w(X^{(s)};\B) = 0$. Fortunately, there exists a simple criterion for identifying such samples without having to compute their permutation numbers explicitly. The criterion comes from Proposition 2.2 in \citet{christensen2023inference}, and asserts that $w(X;\B) > 0$ if and only if $X_{\mathrm{os}}\in\B$, where $X_{\mathrm{os}}$ are the order statistics of $X$. Thus, at each sample $X^{(s)}$ used by the estimator, we first sort the $X_i^{(s)}$ and check whether $X_{\mathrm{os}}^{(s)}\in\B$. If this is the case, we proceed to compute the permutation number exactly. If not, then we discard the sample and continue to the next sample.

\subsection{Logarithmic scaling}
Permanents and subpermanents are often large quantities, and we find it necessary to handle them on a logarithmic scale to avoid integer overflow. More specifically, we only store the logarithms of subpermanents $\per_{r,s}(A)$ in memory. Recall that the permanent of a matrix $A$ equals the sum of its subpermanents. To compute $\log \per\,A$ without incurring an overflow, we use the \code{LogSumExp} trick: 
$$
\log \per\,A = p^* + \log \left( \sum_{r,s} \exp \left\{ \log \per_{r,s}(A) - p^* \right\} \right),
$$
where $p^* = \underset{r,s}{\max} \ \log \per_{r,s}(A)$ is the largest log subpermanent of $A$. Subtracting $p^*$ from the exponent in the sum ensures that the terms are computable without incurring overflow. 

\subsection{Exploiting sparse data structures}
The most computationally intensive part of \code{get_log_perms} is the iterative procedure in which the reduction from Section \ref{sec:reduction31} is reversed via the recurrence relations from \citet[supplementary material]{christensen2023inference}. Indeed, at each step of this procedure, we need to compute all subpermanents of the relevant matrix. For example, the reduction shown in Figure \ref{fig:reduction} involves nine matrices, and all subpermanents of these have to be iteratively computed. To reduce this computational burden, we make the following key observation: Consider the reduced matrix $A'$. The collection of subpermanents $\{\per_{r,s}(A')\}_{r,s}$ is often \textit{sparse}, meaning that a majority of the elements are zero. Indeed, we saw that $\per_{r,s}(A_1') = 0$ for all but one choice of $r$ and $s$ for the matrix $A_1'$ defined in \eqref{eq:favouritereducedfully}. Consequently, it is undesirable to store these subpermanents in matrix form due to cache inefficiency and excessive memory consumption. Rather, we store the subpermanents of $A'$ in a dictionary, retaining only the nonzero log subpermanents in memory. Dictionaries are unfortunately not included in the standard C library, and we have therefore developed our own dictionary implementation. Our implementation uses the hashing function \texttt{xxHash} \citep{xxhash} in conjunction with linear probing. 

\subsection{Memory recycling}
The function \code{get_log_perms} is highly iterative. Indeed, it computes the permutation number $w(X^{(s)},\mathcal{B})$ for each of the $S$ samples contained in \code{X} sequentially and independently. It is therefore time saving to recycle memory. That is, to use the same memory locations at each iteration. 

In \texttt{perms}, the arrays \code{alpha}, \code{beta}, \code{gamma} representing the binary matrix $A$ are initialised only once. Whenever we move on to a new sample or reduce $A$, these vectors are adjusted accordingly. To store the subpermanents, we alternate between two dictionaries, \code{dict1} and \code{dict2}. The alternating pattern functions as follows: Say we have fully reduced the matrix $A$ in $t$ reduction steps, leading to the matrix $A' = A_t$. The subpermanents of $A_t$ are stored in \code{dict1}, and they are then used to compute and save the subpermanents of $A_{t-1}$, to be stored in \code{dict2}. Since we do not need the subpermanents of $A_t$ when computing those of $A_{t-2}$, we repurpose \code{dict1} to store the subpermanents of $A_{t-2}$. This iterative procedure continues, ultimately resulting in the subpermanents of $A$ being stored in either \code{dict1} or \code{dict2}.

\subsection{Parallelisation}\label{parallelsec}
The computations of the permutation numbers $w(X^{(s)},\mathcal{B})$ are performed independently between the samples $X^{(s)}$. Thus, the code is easily parallelisable, and the R version of \texttt{perms} runs in parallel by specifying \code{parallel = TRUE} (see Section~\ref{rinfosec}). The user can easily parallelise the Python version of \texttt{perms} by splitting the samples $X^{(s)}$ into $T$ batches and using for instance \texttt{joblib} \citep{joblib_developers_joblib_nodate} to create $T$ processes applying the function \code{get_log_perms} on each batch. 

\section{Results in speed gain}\label{sec:results}
In this section we cover the results of a simulation study comparing the speed of \texttt{perms} to that of the original implementation in \citet{christensen2023inference}. Specifically, we consider the average computing time for computing a single (non-zero) permanent for different values of the sample size $n$. We use one of \citeauthor{christensen2023inference}'s toy problems as our setup. For $n=10, 20, \dots, 2000$, we let $0=t_1 < \cdots < t_n = 1$ be uniformly spaced on the unit interval. Next, we generated $U_1, \dots, U_n \overset{\mathrm{iid}}\sim \mathrm{Uniform}[0,1]$ and let $Y_i = \one\{U_i \leq t_i\}$. We then generated $S=10,000$ independent random samples $X^{(s)}$ of size $n$ by letting
$$X_1^{(s)}, \dots, X_n^{(s)} \overset{\mathrm{iid}}\sim \mathrm{Beta}(2,2).$$
We then computed the associated permanents using the old and new methods, respectively, and averaged the run times. To make the simulations run in reasonable time, we only ran the old method up to $n=500$ and averaged over $S'=3000$ samples. Note that when computing these averages, we divide by the number of permanents computed at each step (which changes with $n$), not by $S$ or $S'$. Hence, for each value of $n$, we record the average computing time for a single non-zero permanent for both the old and new method. 
%Note that when computing these averages, we have to divide by the number of permanents computed at each step (which changes with $n$), not by $S$ or $S'$.

\begin{figure}
    \centering
    \includegraphics{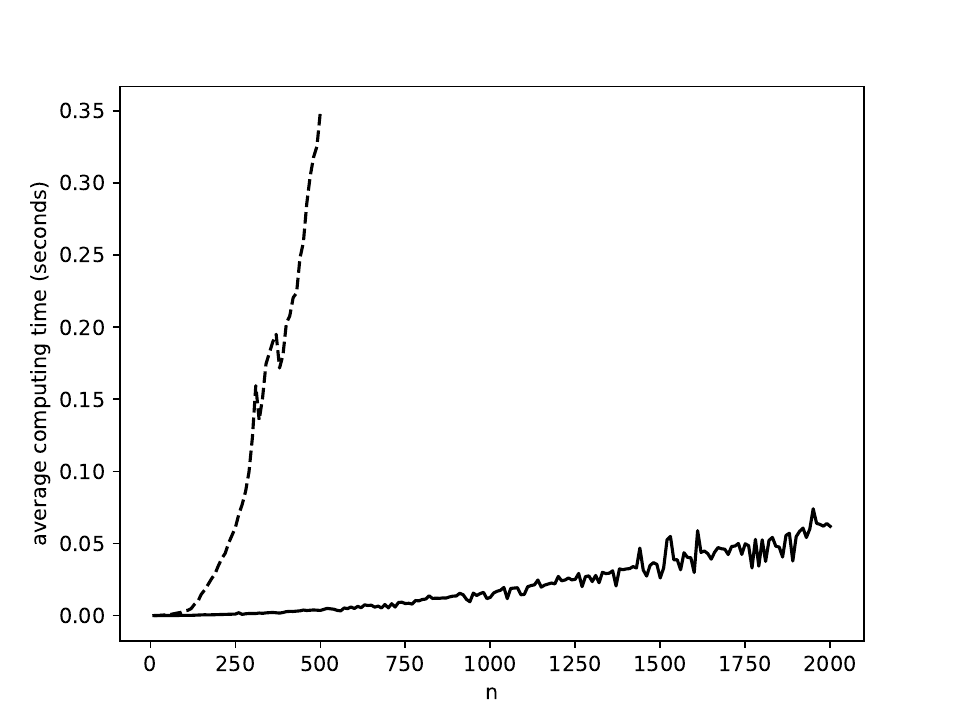}
    \caption{\label{fig:simulationstudy} Run times of the old method (dashed curve) and \texttt{perms} (solid curve) in seconds.}
\end{figure}

\begin{figure}
    \centering
    \includegraphics{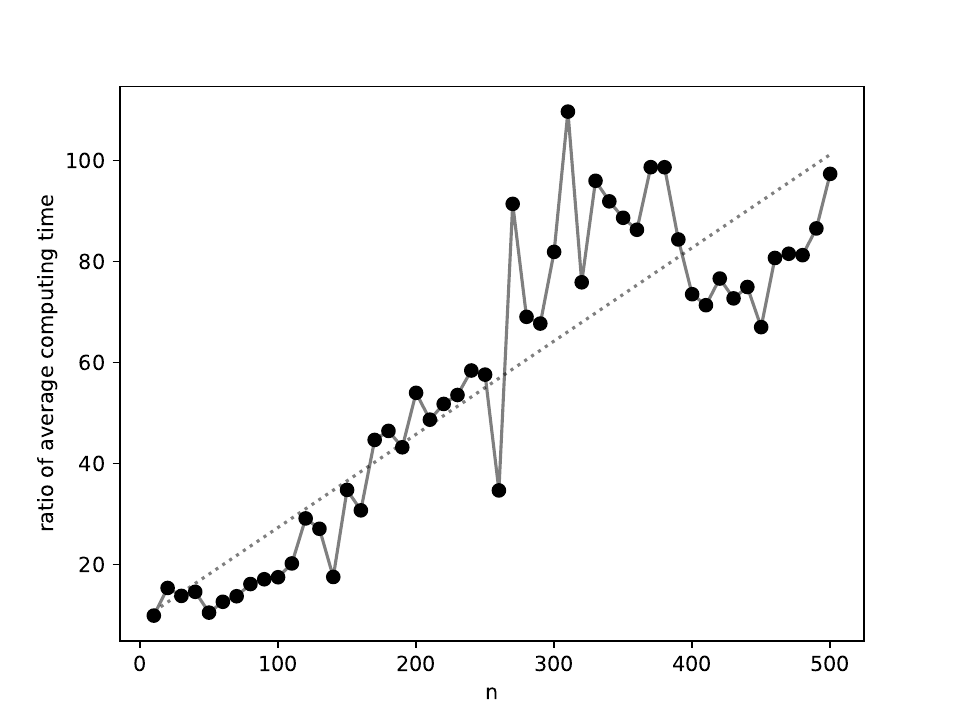}
    \caption{\label{fig:simulationratio} Ratio of average computing times for the old method and the new method, along with the linear fit (dashed line).}
\end{figure}

The results of the simulations are shown in Figure~\ref{fig:simulationstudy}. As we can see, \texttt{perms} outperforms the old implementation drastically. In Figure~\ref{fig:simulationratio} we see the ratio of the average computing times of a non-zero permanent for the old and new methods, respectively. This ratio is increasing with $n$. We get a decent linear trend ($R^2 = 0.79$), indicating that \texttt{perms} is faster than the old method by a factor of $n$.

Regressing the square root of the average run times of \texttt{perms} onto $n$ gives an excellent linear fit ($R^2 = 0.98$), indicating that the algorithm has computational complexity $\mathcal{O}(n^2)$ for computing a single permanent. From the estimated coefficients, we derive the heuristic
$$\mathrm{average\,\,computing\,\,time} \approx Cn^2,$$
measured in seconds, where $C=1.54\times 10^{-8}$. However, this relationship depends on both the model choice and hardware used. It should also be noted that this simulation study does not address the acceptance rate of the estimator \eqref{eq:logMLest}, i.e., the probability that $w(X;\B) > 0$. For a simulation study investigating this, see \citet{christensen2023inference}.

\section{A further example: Nonparametric changepoint detection}\label{sec:examplecp}
We will end by considering a more complicated example than direct computation of marginal likelihoods, like we did in the problems in Section~\ref{sec:basicusage}. In this example we use the Python version of \texttt{perms}. The problem is to detect a \emph{changepoint} in a sequence of binary response data. Such data arise in the setting of measuring the impact sensitivity of explosives. In practice, the data are obtained by repeatedly dropping a weight (typically of mass 2 kg) upon samples of the explosive in question, and observing either an explosion (success) or no explosion (failure) \citep{christensen2023improved}. It has recently been documented that the impact sensitivity of explosive remnants of war changes over time \citep{novik2022analysis, novik2023increased}, motivating a changepoint model. Changepoint detection for parametric models have been studied extensively \citep{fearnhead2006exact, runge2023gfpop, moen2023efficient}. However, significantly less attention has been devoted to nonparametric changepoint analysis, as it is inherently more difficult. Here, we will employ a nonparametric quantile pyramid prior, introduced by \citet{hjort2009quantile}, on the distribution $P$.

\begin{figure}
    \centering
    \includegraphics{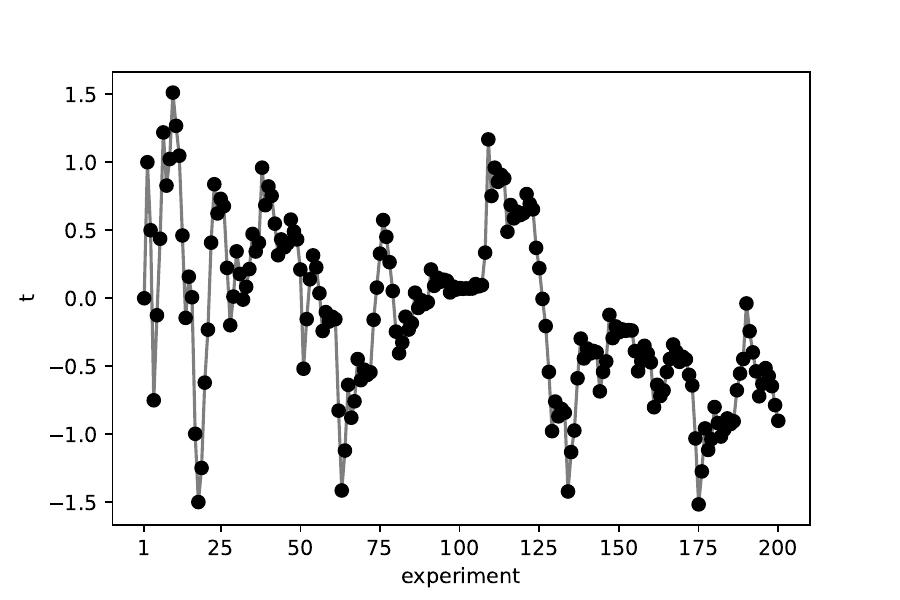}
    \caption{\label{fig:langlie} Inputs $t$ used in the $n=200$ experiments in the changepoint analysis.}
\end{figure}

The data in this example were generated as follows. We simulated $n=200$ experiments with binary outputs $y = (y_1, \dots, y_n)$ according to a probit model. For the first $n_1=120$ experiments, the mean and standard deviation of the probit model were $\mu_1=0.2$ and $\sigma_1 = 1$, whereas for the last $n_2=80$ experiments we used $\mu_2 = -0.7$ and $\sigma_2 = 0.7$. Hence, there is a changepoint at $\tau=120$. The inputs $t = (t_1, \dots, t_n)$ in the experiments were decided by the \citeauthor{langlie1965reliability} test (\citeyear{langlie1965reliability}) on the interval $[-2, 2]$, which is commonly used in the energetic materials industry \citep{baker2021overview}. In Figure~\ref{fig:langlie}, we see the inputs $t$ used for the experiments. We remark that these $t_i$ are treated as fixed covariates, so that the change in distribution at time $\tau$ refers to how the responses $y_i$ relate to the $t_i$. Thus, we are not considering a changepoint in the $t_i$ per se. 

We set the prior probability for the existence of a changepoint to $1/2$. Given that it exists, we use a uniform prior for the changepoint $\tau$ on the domain $\{\gamma, \gamma+1, \dots, n-\gamma-1, n-\gamma\}$ with $\gamma=5$. The domain is cut off on both sides to compensate for the automatic preference for changepoints at the very beginning or very end (see e.g., \citet{cunen2018confidence}). 

In the notation of \citet{hjort2009quantile}, we use a Beta quantile pyramid with parameters $( a_m/2, a_m/2)$, denoted as BetaQP$(a_m/2, a_m/2)$, as prior on the distribution $P$, where $a_m =cm^3$ and $c=2.5$. Since this will generate distributions $P$ with domain $[0,1]$, we apply the transformation $x\mapsto 4x - 2$ to all random samples generated from $P$ to ensure that they have domain $[-2, 2]$ and zero expectation. The null model (no changepoint) and the changepoint model are summarised in Table \ref{tab:changepoint}.

\begin{center}
\begin{table}
\centering
\begin{tabular}{c|c}
\textbf{Null model} & \textbf{Changepoint model} \\ 
\hline
& $\tau \sim \text{Uniform} \left( \{\gamma, \gamma+1, \dots, n-\gamma-1, n-\gamma\}\right)$ \\
& \vspace{-1ex} \\
$P \sim \text{BetaQP}(a_m/2, a_m/2)$ & $P_\ell, P_r \overset{\text{ind}}{\sim} \text{BetaQP}(a_m/2, a_m/2)$  \\
& \vspace{-1ex}\\
$X_1, \ldots, X_n \vert P  \overset{\text{iid}}{\sim} P$ & $\begin{aligned} X_1, \ldots, X_{\tau} \vert \{P_\ell,\tau\}  & \overset{\text{iid}}{\sim}P_{\ell}, \\ X_{\tau+1}, \ldots, X_{n} \vert \{P_r,\tau\}  & \overset{\text{iid}}{\sim}P_r\end{aligned}$  \\
& \vspace{-1ex} \\
$Y_i = \one \{X_i \leq t_i \}$ for $i=1, \dots, n$ & $Y_i = \one \{X_i \leq t_i \}$ for $i=1, \dots, n$. \\
\hline
\end{tabular}
\caption{\label{tab:changepoint} Summary of the null model (left) and changepoint model (right).}
\end{table}
\end{center}

The changepoint problem is twofold. Firstly, we want to assess whether a changepoint exists. Secondly, given that it exists, we want to estimate the location of the changepoint. For the former task, we compute the Bayes factor, which in this case is the ratio of the marginal likelihood of the changepoint model to that of the null model,
\begin{equation*}
    B_{1,0} = \frac{\pi_1(y)}{\pi_0(y)} = \frac1{\pi_0(y)}\sum_{\tau=\gamma}^{n-\gamma}\pi_\ell(y_\ell\mid\tau)\pi_r(y_r\mid\tau)\pi(\tau).
\end{equation*}
Here, $y_\ell = (y_1, \dots, y_\tau)$ and $y_r = (y_{\tau+1}, \dots, y_n)$. For the latter task, we use the posterior mode $\hat\tau$ as our estimate. That is, we find the maximum of the posterior mass function for $\tau$,
\begin{align*}
    \hat\tau & = \argmax_{\tau}\pi(\tau\mid y) \\
    & = \argmax_{\tau}\left\{\log\pi_\ell(y_\ell\mid\tau) + \log\pi_r(y_r\mid\tau)\right\}.
\end{align*}
Thus, both problems require the estimation of marginal likelihoods. The marginal likelihood $\pi_0(y)$ of the null can be estimated directly using \texttt{perms} like we did in Section~\ref{sec:basicusage}. For the marginal likelihood $\pi_1(y)$ of the changepoint model, we use that the latent variables $X_\ell$ and $X_r$ (of $y_\ell$ and $y_r$, respectively) are separately exchangeable. We can therefore estimate $\log\pi_\ell(y_\ell\mid\tau)$ and $\log\pi_r(y_r\mid\tau)$ separately from two matrices \code{X_left} and \code{X_right} of random samples. Note that we do not have to recreate these matrices for each $\tau$. We only have to create them once and then recompute the permanents for each value of $\tau$. In our code, we first create one large matrix \code{X} of dimension \code{(2 * S)} $\times$ \code{n} (with \code{S = 2000}), and then use the first and last \code{S} rows to create \code{X_left} and \code{X_right}, respectively. Having done so, the code for calculating the log marginal likelihood estimate of the null is as follows:

\begin{CodeChunk}
\begin{CodeInput}
log_perms_null = perms.get_log_perms(np.copy(X), t, y, False)
log_ML_null = perms.get_log_ML(log_perms_null, n, False)
\end{CodeInput}
\end{CodeChunk}

Note that we have used a copy of \code{X} in the argument. This is because the \code{get_log_perms} function automatically sorts the rows of \code{X}, but we need the matrix unsorted when computing the marginal likelihood of the changepoint model. To perform this computation, we first define the cutoff \code{gamma = 5} and initialise an array \code{log_MLs} of zeros of length \code{n - 2 * gamma}:

\begin{CodeChunk}
\begin{CodeInput}
gamma = 5
log_MLs = np.zeros(n - 2 * gamma)
\end{CodeInput}
\end{CodeChunk}

We then loop through each candidate changepoint \code{tau}, split the matrix \code{X} into \code{X_left} and \code{X_right} as explained, and then estimate the resulting marginal likelihood using \code{perms}. To do so, we also need to define the corresponding inputs \code{t_left}, \code{t_right} and binary outcomes \code{y_left}, \code{y_left}, respectively. Having calculated the estimates \code{log_ML_left} and \code{log_ML_right}, we sum these and we update the value of \code{log_MLs} at the current index:

\begin{CodeChunk}
\begin{CodeInput}
for tau in range(gamma, n - gamma):
    X_left = X[:S, :tau]
    X_right = X[S:, tau:]
    
    t_left = t[:tau]
    y_left = y[:tau]
    
    t_right = t[tau:]
    y_right = y[tau:]
    
    log_perms_left = perms.get_log_perms(np.copy(X_left), t_left, y_left, 
                                                                   False)
    log_perms_right = perms.get_log_perms(np.copy(X_right), t_right, 
                                                     y_right, False)
    
    log_ML_left = perms.get_log_ML(log_perms_left, tau, False)
    log_ML_right = perms.get_log_ML(log_perms_right, n - tau, False)
    
    log_ML = log_ML_left + log_ML_right
    
    log_MLs[tau - gamma] = log_ML
\end{CodeInput}
\end{CodeChunk}

The computation of all permanents took three minutes and 30 seconds. The Bayes factor can now be computed as follows:

\begin{CodeChunk}
\begin{CodeInput}
    log_Bayes_factor = log_ML - log_ML_null
    Bayes_factor = np.exp(log_Bayes_factor)
\end{CodeInput}
\end{CodeChunk}

\begin{figure}
    \centering
    \includegraphics{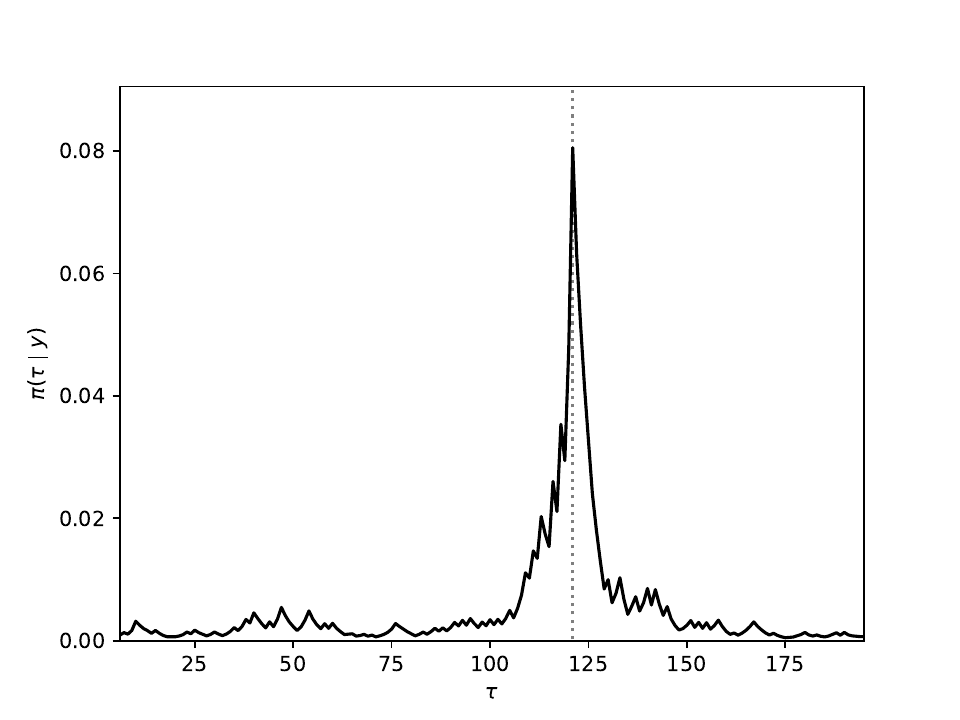}
    \caption{\label{fig:changepointmainfig} Posterior probability mass function of $\tau$ in the changepoint problem, with the posterior mode $\hat\tau = 121$ (dotted line).}
\end{figure}

Doing so, we obtain $B_{1,0} = 4.9571 > 1$, correctly in favour of the changepoint model. Figure~\ref{fig:changepointmainfig} displays the estimated posterior mass function $\pi(\tau\mid y)$. As we can see, the posterior mode is at $\hat\tau=121$, just one off the true value, with high certainty.

\bibliographystyle{plainnat}
\bibliography{refs}

\end{document}